\documentclass[twocolumn,showpacs,preprintnumbers,amsmath,amssymb,floatfix,nofootinbib]{revtex4-1}
\usepackage{color}   %option added by cyw eg \pagecolor{blue} \color{red}
\usepackage{graphicx,float}% Include figure files
\usepackage{dcolumn}% Align table columns on decimal point
\usepackage{bm}% bold math

\usepackage{slashed}

\begin{document}

\onecolumngrid

\twocolumngrid
\def\bb    #1{\hbox{\boldmath${#1}$}}
 \def\oo    #1{{#1}_0 \!\!\!\!\!{}^{{}^{\circ}}~}  
 \def\op    #1{{#1}_0 \!\!\!\!\!{}^{{}^{{}^{\circ}}}~}

\title{ States of the $^{12}$C Nucleus in the  Toroidal Configuration  }

\author{Cheuk-Yin Wong}
\email{wongc@ornl.gov}
\affiliation{Physics Division, 
Oak Ridge National Laboratory, 
Oak Ridge, TN 37831, USA}

\author{Andrzej Staszczak}
\email{andrzej.staszczak@poczta.umcs.lublin.pl}
\affiliation{Institute of Physics, Maria Curie-Sk{\l}odowska University,
pl.~M.~Curie-Sk{\l}odowskiej 1, 20-031 Lublin, Poland}

\date{\today}

\begin{abstract}

The $^{12}$C nucleus with $N$=6 and $Z$=6 is a doubly closed-shell
nucleus in a toroidal potential. 
 In the description of the    ground state and the Hoyle state of $^{12}$C in the
resonating group method or the generator coordinate method,
a superposition of the orientations
of Wheeler's triangular cluster on the cluster plane would
naturally generate an intrinsic  toroidal density.
A toroidal state also has a probability amplitude to overlap with a 
3-alpha cluster, which is the dominant decay mode for
the Hoyle state. For these reasons, we study a toroidal
description of the states of $^{12}$C in the toroidal configuration
both phenomenologically and microscopically. A toroidal $^{12}$C nucleus 
distinguishes itself by toroidal particle-hole multiplet excitations
between one toroidal single-particle shell to another.  From such a
signature and experimental data, we find 
phenomenologically
that the Hoyle state and many
of its higher excited states may be tentatively attributed to those of
the $^{12}$C nucleus in a toroidal configuration. We then study the
$^{12}$C system from a microscopic mean-field approximation using
variational wave functions. We find that the equidensity surfaces of
the $^{12}$C ground state exhibit a dense toroidal core immersed in lower-density 
oblate spheroids in the surface region. Furthermore, there are
prominent toroidal features of the equidensity surfaces for the state at the Hoyle
excitation energy, at which previous cluster model calculations
indicate the presence of a 3-alpha cluster state. A toroidal
coexistence model therefore may emerge to suggest the possibility that
the physical Hoyle state  may have
probability amplitudes to be in the toroidal configuration and the
3-alpha cluster configuration, possessing features of both the
toroidal particle-hole multiplet signature studied here and the
3-alpha cluster decay properties examined elsewhere.

\end{abstract}


\maketitle

\section{Introduction}

The study of the intrinsic structure of the $^{12}$C nucleus has a
long history.  Wheeler proposed an $\alpha$ particle model for
$\alpha$-conjugate nuclei and suggested in 1937 that the $^{12}$C
nucleus may be described as a triangular resonating grouping of three
alpha clusters obeying Bose-Einstein statistics and exchanging
nucleons between them \cite{Whe37,Whe37a}.  Later in 1953 Hoyle
postulated an excited state of $^{12}$C as the doorway for the triple
alpha reaction in nucleosynthesis, in which two alpha particles fuse
into beryllium-8 and then capture a third alpha particle to form the
carbon-12 nucleus \cite{Dun53,Hoy53,Hoy54}.  The excited 0$^+$ state
at 7.654 MeV of $^{12}$C was subsequently identified as the postulated
``Hoyle state" \cite{Coo57}.  Since then, many experimental and
theoretical investigations have been carried out to understand the
properties of the $^{12}$C nucleus.  Recent experimental and
theoretical results and reviews have been presented in
\cite{Bec10,Whe14,Whe15,Zim13,Smi17,Del17,Gai17,Kel17,Kir10,Kir12,Alc12,Bar18,Hor10,Fre17,Des12,
  AR07,AR07a,AR08,AR09,Fel16}.  In addition to Wheeler's $\alpha$ particle
model for the $^{12}$C nucleus, related theoretical models include the
cluster model of Brink \cite{ Bri66}, the cluster model of three
interacting alpha particles \cite{AR07,AR07a,AR08,AR09}, the generator
coordinate method \cite{Hil53,Ueg77,Ueg79,Des12}, the resonating group
method \cite{Kam72,Kam74,Kam81}, a Bose-Einstein condensate-type
cluster model \cite{Hor74,Fun05,Fun09,Toh01}, the (THSR) cluster wave
function model \cite{Toh01}, a Nilsson oblate ellipsoidal model
\cite{Nil55} with a commensurate axis ratio of $a_\rho$:$a_z$=2:1
\cite{Won70}, an algebraic U(7) model with a D$_{3h}$ symmetry
\cite{Bij00,Bij02,Mar14}, an antisymmetric molecular dynamics (AMD) model
\cite{Oer96,Kan98,Kan01,Kan07}, a microscopic fermionic molecular
dynamic (FMD) model \cite{Rot04,Che07,Fel16}, an {\it ab initio} no-core
shell model \cite{Rot11,Bar13}, a no-core simplectic model (NCSpM)
\cite{Dre12}, {\it ab initio} lattice effective field theory (L-EFT)
\cite{Epe11}, energy-density functional mean-field models
\cite{Egi04,Aru05,Rei11,Ebr12,Ich11,Zha15,Afa18,Sta19}, and a rod
model of $^{12}$C excited states \cite{Su72,Ren18,Ina18}.  For a
review of microscopic cluster models, see
Refs.\ \cite{Bec10,Hor10,Fre17,Des12,Fel16}.

We study here the additional toroidal degree of freedom in the
$^{12}$C nucleus for many reasons.  First and foremost is the reason
that the physical nature of many excited states of $^{12}$C has not
been fully understood
\cite{Bec10,Whe14,Whe15,Zim13,Smi17,Del17,Gai17,Kel17,Kir10,Kir12,Alc12,Bar18,Hor10,Fre17,Des12,
  AR07,AR07a,AR08,AR09}.  On the other hand, Wheeler suggested that
under appropriate conditions, a nucleus may assume a toroidal shape
\cite{Whe50,Whe98}.  Referring to toroidal nuclei and black holes,
Wheeler wrote in his autobiography \cite{Whe98}, ``If nuclei could
exist in doughnut shapes, I felt, then, some of them would exist in
such shapes.  If matter could collapse to infinitesimal or even zero
size, then some matter would collapse.  We physicists should think
about where such extreme behavior might occur, and look for it."  A
possible ``extreme behavior" for a toroidal $^{12}$C nucleus might be the
presence of a strong toroidal shell effect, since the $^{12}$C nucleus
with $N$=6 and $Z$=6 is a doubly closed-shell nucleus in the toroidal
potential \cite{Won72,Won73}.  The extrapolation from the toroidal nuclei results in Fig.\ 18 of
Ref.\ \cite{Won73}, using the shell-correction method \cite{Bra73}, points 
to a possible toroidal state in $^{12}$C in the low excitation energy region.  
The ``extreme behavior" of a strong shell effect in conjunction with
the ``extreme behavior" of a large yrast spin alignment lead to the prediction of  a toroidal high-spin isomer
in $^{28}$Si \cite{Sta14},
for which  possible experimental observation  has recently been presented 
 \cite{Cao19}.
Toroidal
high-spin isomers under similar extreme behavior have also been predicted in the
light-mass region from $^{24}$Mg to $^{56}$Ni in non-relativistic and
relativistic mean-field theories
\cite{Zha10,Ich12,Sta14,Ich14,Sta15,Sta15a,Sta16}.  It is interesting
to note that in the description of the ground state and the Hoyle state for the $^{12}$C nucleus, 
a generator coordinate superposition of the orientations
of Wheeler's triangular cluster on the cluster plane would
naturally generate an intrinsic  toroidal density.  A toroidal state also
has a probability amplitude overlap with the state of three alpha
particles, which is the dominant decay mode for the Hoyle state 
 \cite{Kir10,Kir12,Alc12,Kel17,Bar18}.
For all these reasons, it is of interest to follow  Wheeler's advice to think about where such extreme behavior for a toroidal nucleus might occur and look for it 
among 
states in the $^{12}$C nucleus.

The 0$^+$(ground) and 2$^{+}$(4.43 MeV) states of the $^{12}$C nucleus
have been identified as members of the collective rotational band of a
deformed nucleus with the quadruple deformation parameter $\beta_2
=-0.6$ \cite{Gri67}.  The intrinsic density of the  nucleus with such a strongly negative
quadrupole $\beta_2$ may contain both oblate
spheroidal and toroidal equidensity surfaces.  It is therefore useful
to generalize the concept of the toroidal nucleus to include those for
which {\it some} of its equidensity surfaces are toroidal.

In this first exploration of its kind to examine the toroidal degree
of freedom of the $^{12}$C nucleus, it is appropriate to study the
problem from both the phenomenological and microscopic points of view.
In the phenomenological study, we search for the signature of the
$^{12}$C nucleus in a toroidal configuration 
so as to facilitate its identification.  The toroidal
geometry of a toroidal nucleus is associated with distinctive
properties of the single-particle states.  The intrinsic shape of a
toroidal $^{12}$C nucleus shows up as a bunching of single-particle
states into ``$\Lambda$-shells" whose spacing is intimately tied to
the size of the toroidal major radius.  This set of single-particle
shells will generate a distinct pattern of particle-hole multiplet
excitations that may be utilized to reveal the toroidal nature of the
nucleus.  We shall look for the possible occurrence of such a
signature among the excited states of $^{12}$C to explore
phenomenologically whether the Hoyle state at 7.654 MeV may be the
head of the toroidal band of the $^{12}$C nucleus.

In the subsequent microscopic studies, we use the mean-field
approximation and the Skyrme energy density functional to examine the
energy of the system in many different shapes, ranging from the
spherical, prolate spheroid, oblate spheroid, bi-concave disk, and
toroidal density distributions.  The investigation will be carried out
with variational single-particle wave functions characterized by a set
of shape parameters.  The energy as a function of the shape parameters
provide useful information on the landscape of the energy surface and
on the density distributions at different energy landscape points to
facilitate the investigation on toroidal density distributions of the
ground state and the state at the Hoyle excitation energy.

It is worth emphasizing at this point that while Wheeler's concept of
the $^{12}$C nucleus as a triangular cluster of three alpha particles
\cite{Whe37,Whe37a} has been studied extensively in
\cite{Bri66,AR07,AR07a,AR08,AR09,Hil53,Ueg77,Ueg79,Kam72,Kam74,Hor74,Fun05,Fun09,
  Des12,Kam81}, Wheeler's other concept of possible toroidal nucleus
\cite{Whe50,Whe98} has up to now not been applied to the $^{12}$C
nucleus, even though $^{12}$C is a doubly closed-shell nucleus in a
toroidal potential \cite{Won72,Won73}.  Wheeler's two different
concepts should play their separate and important roles under
different probes of the nucleus.  In matters of 3$\alpha$ decay and
the escape through the external Coulomb barrier, the 3$\alpha$ cluster
description is clearly the simpler description.  However, because 
 nucleons can traverse azimuthal orbitals in a
torus with low energies and can be excited to higher orbitals with a low excitation energies,  a description in terms of a
$^{12}$C nucleus in toroidal doubly-closed shells may be an efficient
description in matters associated with particle-hole excitations and
in the energy spacing between particle-hole multiplet states.  In
contrast, the 3$\alpha$ cluster description has an unperturbed (one
particle)-(one hole) energy excitation energy of order 20 MeV, and a
large mixing of the many 3$\alpha$ determinants will be needed to
bring the unperturbed levels down to the observed particle-hole
excitation energy of a few MeV to about 10 MeV.  Furthermore, the
toroidal concept provides a novel geometrical insight, organizes
useful concepts, helps guide our intuition, and may find many
applications involving the $^{12}$C nucleus.  It is therefore
beneficial to develop the toroidal concepts for the $^{12}$C nucleus
proposed here.

The paper is organized as follows.  In Section II, we write down the
approximate single-particle energies in a light toroidal nucleus which
depend only on the effective toroidal major radius $R$ and the orbital
angular momentum component $\Lambda_z$ along the symmetry $z$-axis.
The method to calculate the quantum numbers and the excitation
energies of the particle-hole excitations from these single-particle
states are examined in Section III.  The spectrum of $^{12}$C in the
toroidal configuration is investigated in Section IV.  In Section V,
we compare the theoretical predicted spectrum of toroidal states with
observed $^{12}$C states.  In Section VI, we propose a toroidal
constraint in mean-field dynamics to study the energy surface in the
toroidal degree of freedom and to locate local energy minima in the
toroidal configuration.  In Section VII, we study the states of
$^{12}$C from a microscopic viewpoint using variational wave functions
in terms of geometrical parameters.  The search for the minimum of the
energy using the Skyrme SkM* interactions \cite{Vau72,Vau73,Bar82}
allows the determination of the ground state energy and density in
Section VIII.  We calculate the adiabatic energy surface and the
density distributions for states above the ground state in Section IX.
We examine the connection between the toroid and three-alpha cluster
configurations.  We study how a phenomenological toroidal coexistence
model may emerge to contemplate the mixing of the toroidal and the
three-alpha cluster descriptions of the $^{12}$C states in Section X.
In Section XI, we present our summary and discussions.

\section{Lowest Single-Particle States in a Toroidal Potential } 

We consider a toroidal nucleus with an axially symmetric density
distribution and choose the symmetry axis to be the $z$-axis.  The
nuclear density will generate an axially symmetric single-particle
mean-field potential $V(\rho,z)$ that is independent of the azimuthal
angle $\phi$.  The single-particle wave function of the lowest states
for a proton or a neutron can be written as
\begin{eqnarray}
\Psi_{n_\rho n_z \Lambda_z \Omega_z}
=\psi_{n_\rho n_z}(\rho,z)
\frac{ [e^{i\Lambda_z\phi}\chi_{s_z}]^{\Omega_z}}{\sqrt{2\pi}},
\end{eqnarray}
and the single-particle wave equation is
\begin{eqnarray}
&&\hspace*{-0.5cm}\biggl [- \frac {\hbar^2}{2m}\left \{ \frac{1}{\rho}\frac {\partial}{\partial \rho}\rho \frac {\partial}{\partial \rho}+\frac{\partial^2}{\partial z^2}-\frac{\Lambda_z^2}{\rho^2} 
 \right \} ~~~~~~~~~~~~~~~
\nonumber\\
&& ~~~~~~~~~~~
 +V(\rho,z) - \epsilon_ {n_\rho n_z \Lambda_z \Omega_z}\biggr  ] \psi_{n_\rho n_z}(\rho,z)
=0,
\label{eq}
\end{eqnarray}
where $n_\rho$ and $n_z$ are the quantum numbers associated with the
oscillations in the $\rho$ and $z$ directions, $\Lambda_z$ is the
orbital angular momentum component along the the $z$-direction, and
$\Omega_z$=$\Lambda_z$+$s_z$.  We are interested in 
the phenomenology of 
a $^{12}$C nucleus with a toroidal density
distribution in $^{12}$C for which the oscillation energy in the $\rho$ and $z$
directions are  substantially greater than the orbital
energy associated with the $\Lambda_z$ quantum number.  In such a case,  the
latter can be treated as a perturbation.  We are also interested in 
particle-hole excitations involving single-particle states with
$n_\rho$=0 and $n_z$=0 whose labels will be omitted.  Upon neglecting
the small spin-orbit interaction for low-lying states with low orbital
angular momenta, the single-particle energy of the lowest states are
\begin{eqnarray}
\epsilon_{\Lambda_z \Omega_z} =
\epsilon_{\rho z 0}  +\frac{\hbar^2 \Lambda_z^2}{2mR^2},
\label{eq3}
\end{eqnarray}
where $\epsilon_{\rho z 0}$ is the zero-point energy associated with
oscillations in the $V(\rho,z)$ potential, and the effective major
radius $R$ is a measure of the matrix element
\begin{eqnarray}
\frac{\hbar^2 \Lambda_z^2 }{2m R^2}=\langle \psi_{n_\rho=0,n_z=0}(\rho,z) | \frac{\hbar^2 \Lambda_z^2}{2m \rho^2} |\psi_{n_\rho=0,n_z=0}(\rho,z) \rangle.~~ 
\end{eqnarray}
To get the lowest lying particle-hole excitation spectrum of the
toroidal nucleus in question, we place the nucleons in the lowest
single-particle states in the toroidal potential, make the
single-particle particle-hole excitations between different
$|\Lambda_z \Omega_z\rangle$ states without exciting the oscillations
in the $\rho$ and $z$ directions, and record their excitation energies
and quantum numbers.

\section{Particle-hole excitations of a Toroidal Nucleus }

With the single-particle state energies given by Eq.\ (\ref{eq3}) for
a toroidal nucleus, the density of single-particle states in the
energy spacing is far from being uniform in a toroidal nucleus.  The
single-particle state energies depend only on the absolute value
$\Lambda$$\equiv$$|\Lambda_z|$.  They group together into toroidal
``$\Lambda$-shells", with a large energy gap between one toroidal
$\Lambda$-shell and the next, generating magic numbers
$N$=$2(2\Lambda+1)$ in the light-mass region \cite{Won72,Won73}.  As
shown schematically in Fig.\ 1, the toroidal ground state of the
$^{12}$C nucleus with the closed toroidal shells is described by
nucleons occupying the $\Lambda$=0 and $\Lambda$=1 toroidal shells,
filling up the lowest single-particle states of $|0
\,\pm\frac{1}{2}\rangle$, $|1 \,\pm\frac{1}{2}\rangle$,
$|1$\,$\pm\frac{3}{2}\rangle$.

\begin{figure} [h]
\includegraphics[scale=0.4]{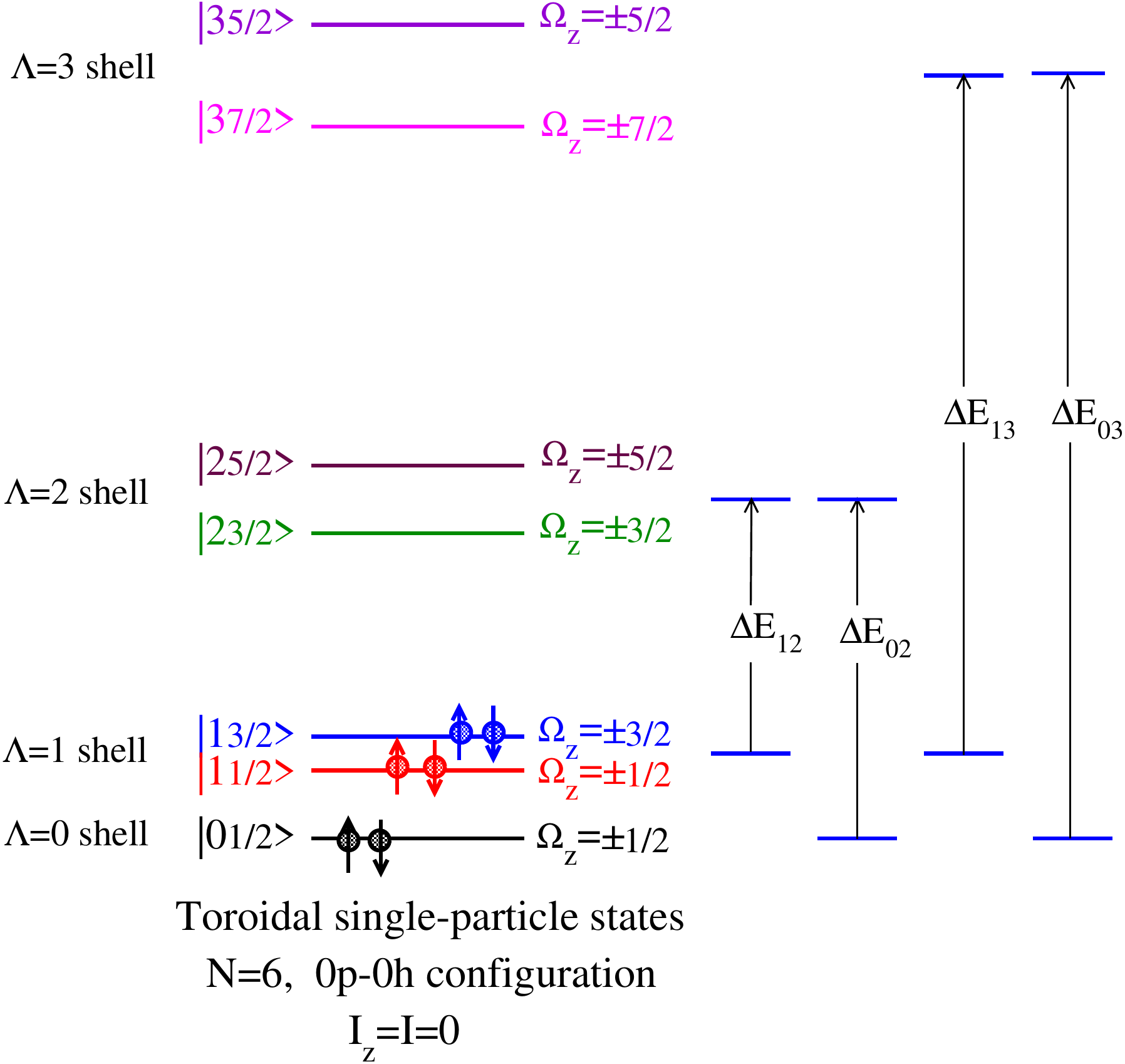} 
\vspace*{0.0cm}
\caption{(Color online.)  Schematic toroidal single-particle energy
  level diagram for neutrons or protons, with the nucleons occupying
  the lowest $|\Lambda_z \Omega_z\rangle$ single-particle states for the $^{12}$C nucleus in the
  toroidal configuration.  The excitation energies $\Delta
  E_{\Lambda_{i} \Lambda_{f}} $ for various (1p1h)$_{\Lambda_i \Lambda_f}$ excitations from
  the $\Lambda_{i}$ toroidal shell to the $ \Lambda_{f} $ toroidal
  shell are schematically indicated.}
\label{fig1} 
\end{figure}

The doubly closed-shell nature of the $^{12}$C nucleus in the toroidal
configuration means that the energy gap between the occupied states in
the $\Lambda$=1 shell and the unoccupied states in the $\Lambda$=2
shell is expected to be quite large, presumably much larger than the
spin-orbit and residual interactions.  Thus, the 
particle-hole excitations from one $\Lambda$-shell to another will yield the gross structure, while the
spin-orbit and residual interactions will provide the fine structure
of the spectrum.  In the present first survey we shall contend
ourselves only with the gross structure by studying particle-hole
excitations built on toroidal single-particle shells without
spin-orbit and residual interactions.  Refinement of the energy
spectrum can be carried out in the projected shell model calculations
using the toroidal states as basis states as in
\cite{Har95,Sun03,Sun04,Sun08}.

We can use the single-particle energies Eq.\ (\ref{eq3}) to 
determine the gross structure of the particle-hole excitation of
a double-closed shell toroidal nucleus by calculating the
particle-hole excitation energy $E_I-E_0$ and its associated
angular momentum component $I_z$. We call the angular
momentum component $I_z$ along the symmetry axis with
$I_z$=$I$ the spin $I$ (or $I_z$) of the
toroidal state.

We construct the multiplet of (n particle)-(n hole) excitations by
promoting n nucleons from the occupied $\Lambda_{i}$-shell to the
empty $\Lambda_{f}$-shell. We label the multiplet as
(npnh)$_{\Lambda_{i}\Lambda_{f}}^\pi$, where $\pi$, the parity of the
multiplet, is equal to $(-1)^{\Lambda_{f}-\Lambda_{i}}$.  In
particular, the (1p1h)$_{\Lambda_{i} \Lambda_{f}}^\pi$ multiplet of
particle-hole excitations can be constructed by promoting a nucleon
from an occupied initial state $| \Lambda_z \Omega_z\rangle_i$ in the
$\Lambda_{i}$-shell to an unoccupied final state $|\Lambda_z
\Omega_z\rangle_f$ in the $\Lambda_{f}$-shell.  Such an $\{i\to f\}$ particle-hole
excitation leads to a spin increment $\Delta_{if} I_z$ given by
\begin{eqnarray}
 \Delta_{if} I_z([\Lambda_z \Omega_z]_i \!\to\!\!  [\Lambda_z \Omega_z]_f\!) =\Omega_{zf} - \Omega_{zi},
\end{eqnarray}
and from Eq.\ (\ref{eq3})
an excitation energy increment $\Delta_{if}  E_{\Lambda_i \Lambda_f}$ given by 
\begin{eqnarray}
\Delta_{if} E_{\Lambda_i \Lambda_f}([ \Lambda_z \Omega_z]_i\!\to\!\!   [ \Lambda_z \Omega_z]_f) 
&&=\!
 \epsilon([\Lambda_z \Omega_z]_f)\!-\! \epsilon([ \Lambda_z \Omega_z]_i)
 \nonumber \\
&&=
 \frac{\hbar ^2}{2m R^2} \left ( \Lambda_{f}^2-\Lambda_{i}^2\right ).~~~~
\label{eq6}
\end{eqnarray}
Here the energy unit $\hbar ^2/2m R^2$ appears so frequently in the
excitation energy expressions that it deserves a symbol of its own.
We call it $\epsilon_0$,
\begin{eqnarray}
\epsilon_0=\frac{\hbar^2}{2m R^2}.
\end{eqnarray}

The spin $I_z$ and excitation energy $E_x$ of a state in the
(npnh)$_{\Lambda_{i}\Lambda_{f}}^\pi$ multiplet, in the simplest
approximation with the neglect of spin-orbit and residual
interactions, are just additive sums of the spin and energy increments
from independent $\{i$$\to$$f\}$ particle-hole excitations, subject to
the Pauli exclusion principle,
\begin{eqnarray}
I_z=\sum_{if} \Delta_{if} ([\Lambda_z \Omega_z]_i \!\to\!\!  [\Lambda_z \Omega_z]_f\!).
\end{eqnarray}
The parity $\pi$ of the state is
\begin{eqnarray}
\pi=\prod_{if} \left \{ (-1)^{\Lambda_{f}-\Lambda_{i} }\right \}.
\end{eqnarray}
Because there are many different $\Omega_z$ states in a
single-particle $\Lambda$-shell, there are many different spin
increments in an (npnh)$_{\Lambda_{i}\Lambda_{f}}^\pi$ multiplet.
Consequently there are many different $I_z$ spins in the multiplet
states of the toroidal nucleus.  All these states with different spins
$I=I_z$ within the multiplet have the same parity $\pi$ and are
degenerate with the excitation energy
\begin{eqnarray}
E_x = E_I-E_0=\sum_{if} \Delta_{if} E_{\Lambda_i \Lambda_f}([ \Lambda_z \Omega_z]_i\!\to\!\!   [ \Lambda_z \Omega_z]_f) ,~~~~~~
\label{eq15}
\end{eqnarray}
where $E_0$ is the energy of the toroidal ground state relative to the
$^{12}$C ground state.  The inclusion of spin-orbit and residual
interactions in a more refined calculation will split the degeneracy
of the different $I_z$ states in the multiplet.

\section{Spectrum of the $^{12}$C nucleus in the Toroidal  Configuration }

\subsection{ (1{\MakeTextLowercase p}1{\MakeTextLowercase h})$_{\Lambda_{i}\Lambda_{f}}^\pi$ Toroidal States}

As one can see from Fig.\ \ref{fig1}, different
(1p1h)$_{\Lambda_{i}\Lambda_{f}}^\pi$ multiplets of exited $^{12}$C
toroidal states arise by promoting nucleons from $\{[ 0(\pm 1/2) ]$,$[
  1 (\pm 1/2) ]$,$[ 1 (\pm 3/2) ]\}$ toroidal states in the
$\Lambda$=0 and 1 shells to occupy empty $\{[2 (\pm 5/2)]$,$[ 2 (\pm
  3/2) ]$,$[3 (\pm 7/2)]$,$[3 (\pm 5/2)]\}$ toroidal states in the
$\Lambda$=2 and 3 shells.  The knowledge of the particle and hole
state quantum numbers gives the spin and parity of the excitation, and
the excitation energy can be determined from the single-particle
energy (\ref{eq3}) or the energy increment (\ref{eq6}).  In
particular, for a member $I_z^\pi$ with $I=I_z$ in the
(1p1h)$_{\Lambda_{i} \Lambda_{f}}^\pi$ multiplet, the energy $E_I$ of
the member relative to the toroidal ground state of energy $E_0$ is
$
E_I-E_0=\Delta E_{\Lambda_{i} \Lambda_{f}}$.
All members of a multiplet have the same excitation energy in the
present idealized approximation with the neglect of the spin-orbit and
residual interactions.

The  (one particle)-(one hole) 
 $\Delta
E_{\Lambda_{i} \Lambda_{f}}$
for different shell-to-shell  excitations are
\begin{subequations}
\begin{eqnarray}
\Delta E_{12}=\epsilon( \Lambda=1)\to \epsilon(\Lambda=2)&& = 
 \frac{3\hbar ^2}{2m R^2} =  3\epsilon_0,
\label{eq11a}
 \\
 \Delta E_{02}=\epsilon( \Lambda=0) \to \epsilon(\Lambda=2)&&= 
 \frac{4\hbar ^2}{2mR^2}= 4 \epsilon_0,
 \\
 \Delta E_{13}= \epsilon( \Lambda=1)\to \epsilon(\Lambda=3)&& = 
 \frac{8\hbar ^2}{2mR^2}=8 \epsilon_0,~~~
\\
 \Delta E_{03}= \epsilon( \Lambda=0)\to\epsilon( \Lambda=3)&& = 
 \frac{9\hbar ^2}{2mR^2}=9 \epsilon_0.~~~~~~
\label{eq11d}
\end{eqnarray}
\label{eq11}
\end{subequations}
These excitation energies depend only on a single-parameter, the major
radius $R$, which can be determined by confronting the predicted
theoretical spectrum with the experimental data in the next Section.
For particle-hole  excitations among the lowest states with $n_\rho=0$ and $n_z=0$,
the above equations indicate that the excitation energies are indepedent of the shape of the underlying toroidal potential 
$V(\rho,z)$. 
The parity of a member $I_z^\pi$ member of the (1p1h)$_{\Lambda_{i}
  \Lambda_{f}}^\pi$ multiplet, is
$\pi=(-1)^{\Lambda_{f}-\Lambda_{i}}$.

\begin{table}[H]
\centering
\caption {The  (1p1h)$_{\Lambda_{i} \Lambda_{f}}^\pi$
excitations in toroidal $^{12}$C where $[\Lambda_{f} \Omega_{zf}]$ represents
a particle state and $[\Lambda_{i} \Omega_{zf}]^{-1}$  represents the hole state.}
\begin{tabular}{|c|c|c|c|}
\hline
particle-hole &  particle-hole &  &     \\
excitation  &  configuration  &   $I_z^\pi$ & $\frac{E_I-E_0}{\epsilon_0}$  \\
 \hline
&$[1($-$3/2)]^{-1}$[2(5/2)]  & 4$^{-}$ & \\
 & $[1($-$1/2)]^{-1}$[2(5/2)]  & 3$^{-}$ & \\
&$[1(~1/2)]^{-1}$[2(5/2)]  & 2$^{-}$  & \\
(1p1h)$_{12}^-$ &$[1(~3/2)]^{-1}$[2(5/2)]  & 1$^{-}$  & \\
\cline{2-3}
$\Lambda$=$1$ shell $\to$$\Lambda$=2 shell &$[1($-$3/2)]^{-1}$[2(3/2)]  & 3$^{-}$ &3  \\
 &$[1($-$1/2)]^{-1}$[2(3/2)]  & 2$^{-}$ & \\
&$[1(~1/2)]^{-1}$[2(3/2)]  & 1$^{-}$  &\\
&$[1(~3/2)]^{-1}$[2(3/2)]  & 0$^{-}$  &\\
 \hline
 &$[0($-$1/2)]^{-1}$[2(5/2)]  & 3$^{+}$ &  \\
(1p1h)$_{02}^+$ &$[0(~1/2)]^{-1}$[2(5/2)]  & 2$^{+}$ & \\
\cline{2-3}
$\Lambda$=$0$ shell $\to$ $\Lambda$=2 shell & $[0($-$1/2)]^{-1}$[2(3/2)]  & 2$^{+}$ & 4 \\
& $[0(~1/2)]^{-1}$[2(3/2)]  & 1$^{+}$ &  \\
 \hline
&$[1($-$3/2)]^{-1}$[3(7/2)]  & 5$^{+}$ & \\
&$[1($-$1/2)]^{-1}$[3(7/2)]  & 4$^{+}$ & \\
&$[1(~1/2)]^{-1}$[3(7/2)]  & 3$^{+}$  &\\
(1p1h)$_{13}^+$ &$[1(~3/2)]^{-1}$[3(7/2)]  & 2$^{+}$ & \\
\cline{2-3}
$\Lambda$=$1$ shell $\to$ $\Lambda$=3 shell &$[1($-$3/2)]^{-1}$[3(5/2)]  & 4$^{+}$ &8 \\
&$[1($-$1/2)]^{-1}$[3(5/2)]  & 3$^{+}$& \\
&$[1(~1/2)]^{-1}$[3(5/2)]  & 2$^{+}$  &\\
&$[1(~3/2)]^{-1}$[3(5/2)]  & 1$^{+}$ & \\
\hline
&$[0($-$1/2)]^{-1}$[3(7/2)]  & 4$^{-}$ &  \\
(1p1h)$_{03}^-$&$[0(~1/2)]^{-1}$[3(7/2)]  & 3$^{-}$ & \\
\cline{2-3}
$\Lambda$=$0$ shell $\to$ $\Lambda$=3 shell &$[0($-$1/2)]^{-1}$[3(5/2)]  & 3$^{-}$ & 9\\
&$[0(~1/2)]^{-1}$[3(5/2)]  & 2$^{-}$ & \\
\hline
\end{tabular}
\label{tb3}
\end{table}

We show the spin quantum number $I_z$, parity $\pi$, and the
excitation energy $E_I-$ $E_0$ of different (1p1h)$_{\Lambda_{i}
  \Lambda_{f}}^\pi$ multiplets of excited toroidal states in Table I.
We have kept the sign of $\Omega_z$ of the particle state positive.
If we study the remaining case by reversing the sign of $\Omega_z$ of
the particle state, we obtain the same set of states as in Table I,
except that the signs of $I_z$ is reversed if it is non-zero, and it
has a different particle-hole combination if $I_z$ is zero.  Thus,
each of the total set of states in Table I is doubly degenerate.  The
double degeneracy occurs repeatedly in all toroidal states, and we
shall make its double degeneracy implicit and shall consider only
non-negative values of $I_z$ with double degeneracy in what follows,
unless explicitly specified otherwise.  Furthermore, it should be kept
in mind that Table I is applicable to neutron as well as to proton
(1p1h) excitations.

\begin{figure} [H]
\centering
\includegraphics[scale=0.47]{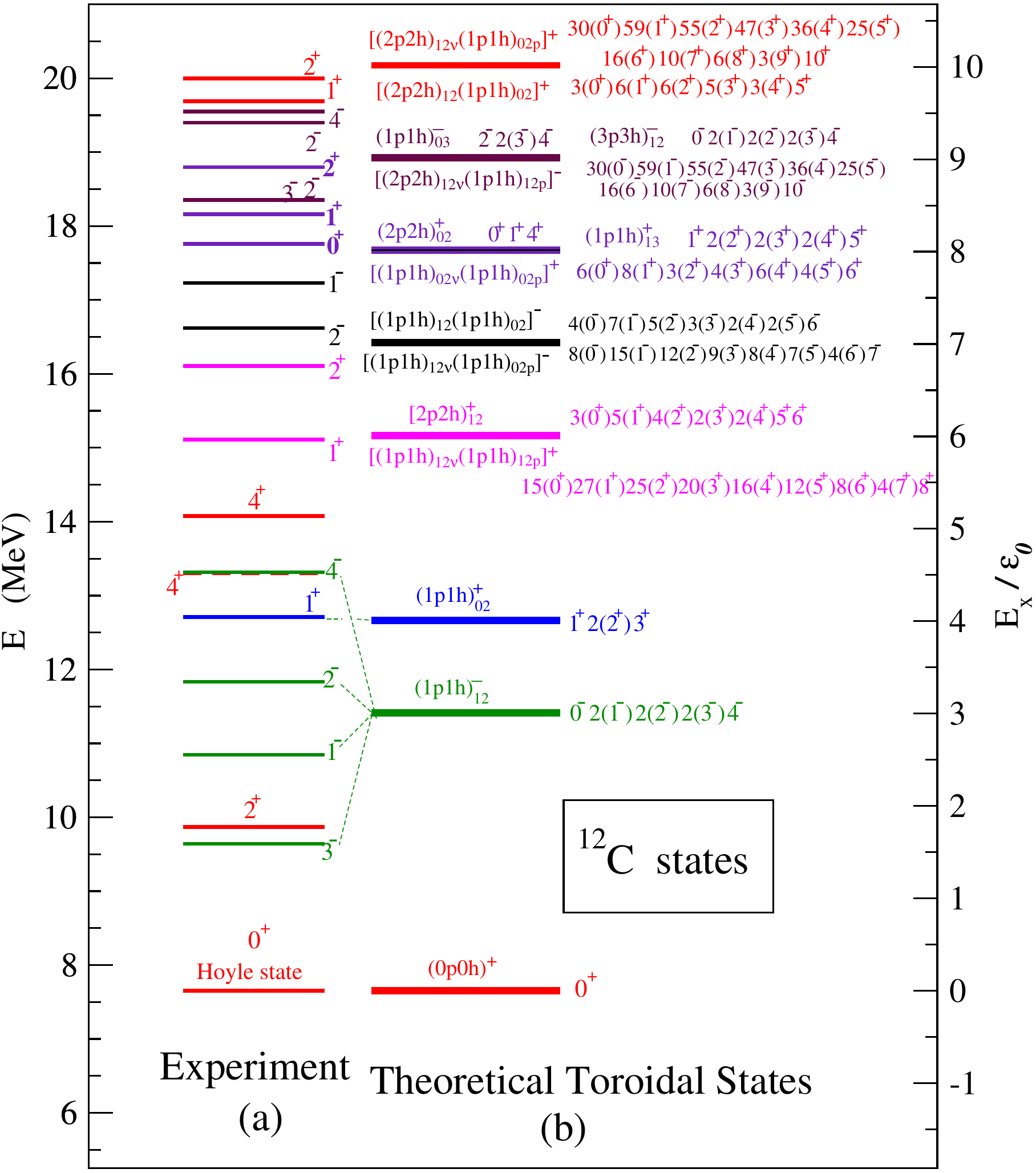} 
\caption{(color online).  (a) Experimental excitation energy $E$ of
  $^{12}$C excited states relative to the $^{12}$C ground state
  \cite{Kel17}, with the axis of the excitation energy $E$ on the
  left.  (b) The theoretical spectrum of the toroidal states in
  different multiplets, with the axis of the excitation energy
  $E_x=E-E_0$, relative to the energy $E_0$ of the toroidal ground
  state, given on the right.  Comparison between the experimental and
  theoretical spectra is made by identifying the Hoyle state as the
  toroidal ground state and the lowest lying $3^-, 1^-,2^-,4^-$ as
  members of the toroidal (1p1h)$_{12}^-$ multiplet, leading to
  $\epsilon_0=1.25$ MeV (see text).  The spins and parities of members
  of the multiplets are presented in the figure and given in Table
  \ref{tb6}.  }
\label{fig2}
\end{figure}

Table I and Fig.\ 2 show that the (1p1h)$_{12}^\pi$ multiplet for the
excitation of a nucleon from the $ \Lambda$=1 shell to the $\Lambda$=2
shell consists of a set of eight doubly-degenerate states,
$\{4^{-},2(3^-),2(2^-),2(1^-)$,$0^-\}$, lying at \break
$E_x$=$E_I$-$E_0 $=$3\epsilon_0$.  The (1p1h)$_{02}^+$ multiplet for
the excitation of a nucleon from the $\Lambda$=0 shell to the
$\Lambda$=2 shell consists of a set of four states,
$\{(3^+),2(2^+),1^+\}$, lying at an excitation energy $E_x
$=$4\epsilon_0$.  The (1p1h)$_{13}^+$ multiplet for the excitation of
a nucleon from the $ \Lambda$=1 shell to the $\Lambda$=3 shell
consists of a set of eight states of
$\{5^+,2(4^+),2(3^+),2(2^+),1^+\}$ lying at $E_x$=8$\epsilon_0$.
Finally, the (1p1h)$_{03}^-$ multiplet for the excitation of a nucleon
from the $\Lambda$=0 shell to the $\Lambda$=3 shell consists of a set
of four doubly-degenerate states, $\{(4^-),2(3^-),2^-\}$, lying at
$E_x $=$ 9\epsilon_0$.  The spectrum of these states are shown in
Fig.\ \ref{fig2}.

The lowest lying toroidal states are the (0p0h)$^+$, (1p1h)$_{12}^-$, and
(1p1h)$_{02}^+$ multiplets lying above the ground toroidal $^{12}$C
state.  They should be prominently excited by stripping ($^3$He,d)
reactions that add a proton to excite and combine with a $^{11}$B
nucleus to a toroidal configuration.

\subsection{ (2p2h)$_{12}^+$  at 
 $E_x$=$6\epsilon_0$ for exciting two identical nucleons 
 from  
$\Lambda$=1 shell to $ \Lambda$=2 shell
}

We consider next the (2p2h)$_{12}^+$ excitations of two identical
particles (neutrons or protons) from the $\Lambda$=1 shell to the
$\Lambda$=2 shell.  Because of the Pauli exclusion principle, the two
identical particle or holes cannot occupy the same
$|\Lambda_z,\Omega_z\rangle$ state.  Consequently, to get the (2
particle)-(2 hole) excitations involving identical particles, it is
simplest to combine the angular momentum components of the two
particles and two holes separately first, under the restriction of the
Pauli principle, before combining them together.  For this purpose, we
list all combinations of states of two holes in Table \ref{tab2h} and
two particles in Table \ref{tab2p}, under the restriction of the Pauli
principle.
\vspace*{0.3cm} 
\begin{table}[h]
\caption {Combination of two holes states in $| 1 (\pm 3/2)\rangle$
  and $| 1 ( \pm 1/2)\rangle$ states in the $\Lambda$=1 shell under
  the restriction of the Pauli principle }
\vspace*{0.3cm} 
\begin{tabular}{|c|c|c|}
\hline
  two hole configuration  &   $I_z^\pi$    \\ \hline
$\{ [1(~3/2)][1(~1/2)]\}^{-1}$  & [2$^{+}$]$^{-1}$  \\
$\{ [1(~3/2)][1($-$1/2)]\}^{-1}$  & [1$^{+}$]$^{-1}$  \\
$\{ [1(~1/2)][1($-$1/2)]\}^{-1}$  & [0$^{+}$]$^{-1}$  \\
$\{ [1($-$3/2)][1(~3/2)]\}^{-1}$  & [0$^{+}$]$^{-1}$  \\
$\{ [1($-$3/2)][1(~1/2)]\}^{-1}$  &[(-1)$^{+}$]$^{-1}$  \\
$\{ [1($-$3/2)][1($-$1/2)]\}^{-1}$  & [(-2$)^{+}$]$^{-1}$  \\
 \hline
\end{tabular}
\label{tab2h}
\end{table}
%\vspace*{0.3cm} 
\begin{table}[H]
\caption {Combination of two particles states in $| 2 \pm 5/2\rangle$
  and $| 2 \pm 3/2\rangle$ orbitals in the $\Lambda$=2 shell under the
  restriction of the Pauli principle } \centering
\vspace*{0.3cm}
\begin{tabular}{|c|c|c|}
\hline
  two particle configuration  &   $I_z^\pi$    \\ \hline
$\{ [2(~5/2)][2(~3/2)]\}$  & 4$^{+}$  \\
$\{ [2(~5/2)][2($-$3/2)]\}$  & 1$^{+}$  \\
$\{ [2(~3/2)][2($-$3/2)]\}$  & 0$^{+}$  \\
$\{ [2($-$5/2)][2(~5/2)]\}$  & 0$^{+}$  \\
$\{ [2($-$5/2)][2(~3/2)]\}$  & (-1)$^{+}$  \\
$\{ [2($-$5/2)][2($-$3/2)]\}$  & (-4)$^{+}$  \\
 \hline
\end{tabular}
\label{tab2p}
\end{table}
Now combine the two hole states with all two particle states, we get
the angular momentum component $I_z$ and parity $\pi$ as listed in
Table \ref{tab2p2h}.  As one observes in Table \ref{tab2p2h}, the
(2p2h)$_{12}^+$ multiplet consists of a set of 18 doubly-degenerate
positive parity states, $\{ 6^+, 5^+, 2(4^+), 2(3^+),4(2^+),5(1^+),
3(0^+)\}$, lying at $E_x=6\epsilon_0$ as shown in Fig.\ \ref{fig2}(b).
It is applicable to a pair of neutrons or protons.

\begin{table}[H]
\caption { The spin states of the (2p2h)$_{12}^+$ multiplet at
  $E_x$=$6\epsilon_0$ for the excitation of two identical nucleons
  from the $ \Lambda$=1 shell to the $ \Lambda$=2 shell }
\vspace*{0.3cm}
\begin{tabular}{|c|c|c|c|}
\hline
two holes &   two particle configuration  &   $I_z^\pi$    \\ \hline
 &   $\{ [2(~5/2)][2(~3/2)]\}$  & 2$^{+}$  \\
 &   $\{ [2(~5/2)][2($-$3/2)]\}$  & (-1)$^{+}$  \\
 &   $\{ [2(~3/2)][2($-$3/2)]\}$  & (-2)$^{+}$  \\
$\{ [1(~3/2)][1(~1/2)]\}^{-1}$  &  $\{ [2($-$5/2)][2(~5/2)]\}$  & (-2)$^{+}$  \\
 &   $\{ [2($-$5/2)][2(~3/2)]\}$  & (-3)$^{+}$  \\
 &   $\{ [2($-$5/2)][2($-$3/2)]\}$  & (-6)$^{+}$  \\
 \hline
  &   $\{ [2(~5/2)][2(~3/2)]\}$  & 3$^{+}$  \\
 &   $\{ [2(~5/2)][2($-$3/2)]\}$  & 0$^{+}$  \\
 &   $\{ [2(~3/2)][2($-$3/2)]\}$  & (-1)$^{+}$  \\
 $\{ [1(~3/2)][1($-$1/2)]\}^{-1}$   &  $\{ [2($-$5/2)][2(~5/2)]\}$  & (-1)$^{+}$  \\
 &   $\{ [2($-$5/2)][2(~3/2)]\}$  & (-2)$^{+}$  \\
 &   $\{ [2($-$5/2)][2($-$3/2)]\}$  & (-5)$^{+}$  \\
 \hline
  &   $\{ [2(~5/2)][2(~3/2)]\}$  & 4$^{+}$  \\
 &   $\{ [2(~5/2)][2($-$3/2)]\}$  & 1$^{+}$  \\
 &   $\{ [2(~3/2)][2($-$3/2)]\}$  & 0$^{+}$  \\
 $\{ [1(~1/2)][1($-$1/2)]\}^{-1}$   &  $\{ [2($-$5/2)][2(~5/2)]\}$  & 0$^{+}$  \\
 &   $\{ [2($-$5/2)][2(~3/2)]\}$  &(-1)$^{+}$  \\
 &   $\{ [2($-$5/2)][2($-$3/2)]\}$  & (-4)$^{+}$  \\
 \hline
  &   $\{ [2(~5/2)][2(~3/2)]\}$  & 4$^{+}$  \\
 &   $\{ [2(~5/2)][2($-$3/2)]\}$  & 1$^{+}$  \\
 &   $\{ [2(~3/2)][2($-$3/2)]\}$  & 0$^{+}$  \\
$\{ [1($-$3/2)][1(~3/2)]\}^{-1}$   &  $\{ [2($-$5/2)][2(~5/2)]\}$  & 0$^{+}$  \\
 &   $\{ [2($-$5/2)][2(~3/2)]\}$  & (-1)$^{+}$  \\
 &   $\{ [2($-$5/2)][2($-$3/2)]\}$  & (-4)$^{+}$  \\
 \hline
  &   $\{ [2(~5/2)][2(~3/2)]\}$  & 5$^{+}$  \\
 &   $\{ [2(~5/2)][2($-$3/2)]\}$  & 2$^{+}$  \\
 &   $\{ [2(~3/2)][2($-$3/2)]\}$  & 1$^{+}$  \\
 $\{ [1($-$3/2)][1(~1/2)]\}^{-1}$   &  $\{ [2($-$5/2)][2(~5/2)]\}$  & 1$^{+}$  \\
 &   $\{ [2($-$5/2)][2(~3/2)]\}$  & 0$^{+}$  \\
 &   $\{ [2($-$5/2)][2($-$3/2)]\}$  & (-3)$^{+}$  \\
\hline
 &   $\{ [2(~5/2)][2(~3/2)]\}$  & 6$^{+}$  \\
 &   $\{ [2(~5/2)][2($-$3/2)]\}$  & 3$^{+}$  \\
 &   $\{ [2(~3/2)][2($-$3/2)]\}$  & 2$^{+}$  \\
 $\{ [1($-$3/2)][1($-$1/2)]\}^{-1}$   &  $\{ [2($-$5/2)][2(~5/2)]\}$  & 2$^{+}$  \\
 &   $\{ [2($-$5/2)][2(~3/2)]\}$  & 1$^{+}$  \\
 &   $\{ [2($-$5/2)][2($-$3/2)]\}$  &(-2)$^{+}$  \\
\hline
\end{tabular}
\label{tab2p2h}
\end{table}

\subsection{ $[$(1p1h)$_{12\nu}$(1p1h)$_{12p}]^+$  toroidal multiplet
 involving  one neutron and one proton } 

In the last subsection, we have considered the (2p2h) excitations
involving two identical nucleons.  The case of the (2p2h) excitations
involving two different types of nucleons from the $\Lambda=1$ shell
to the $\Lambda=2$ shell differ from the previous case, because the
pairs of particles or holes do not need to be restricted by the Pauli
principle.  We can consider such (2p2h) excitations by combining a
(1p1h) neutron excitation with an independent (1p1h) proton
excitation.  Such a [(1p1h)$_{12\nu}$(1p1h)$_{12p}]^\pi$ multiplet
has the excitation energy $E_x=2\Delta E_{12}$=$6 \epsilon_0$ and
positive parity.  In the notation for the multiplet, we have used the subscript $\nu$ for neutrons and $p$ for protons.
Their spin quantum numbers $I_z$ and parities are
given in Table \ref{tb5}.

\vspace*{0.3cm} 
\begin{table}[H]
\caption {  The spin states of the [(1p1h)$_{12 \nu}$(1p1h)$_{12p}]^+$ multiplet at 
 $E_x$=$6\epsilon_0$
involving different types of nucleons }
\vspace*{0.3cm}
\begin{tabular}{|c|c|c|c|c|c|c|c|c|}
\cline{2-9}
 \multicolumn{1}{c|} {}  &    \multicolumn{7}{c} { ~~~~(1p1h)$_{12p}$   State}& \\ 
%\cline{2-9}
\hline
  (1p1h)$_{12\nu}$ &  &  &  & & &  & & \\
 State & 4$^-$ &3$^-$  & 2$^{-}$ &1$^-$&3$^-$&2$^{-}$&1$^-$&0$^-$\\
\hline
  &  & & &  & & & &  \\
  4$^-$  &8$^+$&7$^+$& 6$^+$&5$^+$&
  7$^+$& 6$^+$&5$^+$&4$^+$ \\
\cline{2-9}  
  3$^-$ &7$^+$& 6$^+$ & 5$^+$ & 4$^+$ & 6$^+$ & 5$^+$ & 4$^+$ &3$^+$\\
\cline{2-9}
2$^-$   & 6$^+$ & 5$^+$ & 4$^+$ & 3$^+$ & 5$^+$ & 4$^+$ & 3$^+$ & 2$^+$  \\
\cline{2-9}
1$^-$  &  5$^+$ & 4$^+$ & 3$^+$ & 2$^+$ & 4$^+$ & 3$^+$ & 2$^+$ & 1$^+$  \\
\hline
  3$^-$ &7$^+$& 6$^+$ & 5$^+$ & 4$^+$ & 6$^+$ & 5$^+$ & 4$^+$ &3$^+$\\
\cline{2-9}
2$^-$   & 6$^+$ & 5$^+$ & 4$^+$ & 3$^+$ & 5$^+$ & 4$^+$ & 3$^+$ & 2$^+$  \\
\cline{2-9}
1$^-$  &  5$^+$ & 4$^+$ & 3$^+$ & 2$^+$ & 4$^+$ & 3$^+$ & 2$^+$ & 1$^+$  \\
\cline{2-9}
0$^{-}$ & 4$^+$ &3$^+$  & 2$^+$ &1$^+$&3$^+$&2$^{+}$&1$^+$&0$^+$\\
\hline
\hline
  (-4)$^-$  &0$^+$&(-1)$^+$& (-2)$^+$&(-3)$^+$&
  (-1)$^+$& (-2)$^+$&(-3)$^+$&(-4)$^+$ \\
\cline{2-9} 
   (-3)$^-$ &1$^+$& 0$^+$& (-1)$^+$& (-2)$^+$&0$^+$&
     (-1)$^+$&(-2)$^+$&(-3)$^+$ \\
\cline{2-9} 
(-2)$^-$   & 2$^+$&1$^+$& 0$^+$& (-1)$^+$& 1$^+$&0$^+$&
     (-1)$^+$&(-2)$^+$ \\
\cline{2-9}
(-1)$^-$  &3$^+$& 2$^+$&1$^+$& 0$^+$&  2$^+$&1$^+$& 0$^+$&
     (-1)$^+$ \\
\hline
   (-3)$^-$ &1$^+$& 0$^+$& (-1)$^+$& (-2)$^+$&0$^+$&
     (-1)$^+$&(-2)$^+$&(-3)$^+$ \\
\cline{2-9} 
(-2)$^-$   & 2$^+$&1$^+$& 0$^+$& (-1)$^+$& (-2)$^+$&0$^+$&
     (-1)$^+$&(-2)$^+$ \\
\cline{2-9}
(-1)$^-$  &3$^+$& 2$^+$&1$^+$& 0$^+$&  2$^+$&1$^+$& 0$^+$&
     (-1)$^+$ \\
\cline{2-9}
0$^{-}$ & 4$^+$ &3$^+$  & 2$^+$ &1$^+$&3$^+$&2$^{+}$&1$^+$&0$^+$\\
\hline
\end{tabular}
\label{tb5}
\end{table}

There are altogether $\{$8$^+$,4(7$^+$),8(6$^+$),12(5$^+)$,15(4$^+$),
16(3$^+$),16(2$^+$),15(1$^+$),15(0$^+$),1(-4)$^+$,4(-3)$^+$),9(-2)$^+$),
12(-1)$^+\}$ for a total of 128 states.  If we reverse the sign of the
(1p1h)$_{12p}$ state, we get the same set of states but with the signs
of $I_z$ is reversed if it is non-zero, and it has a different
particle-hole combination if $I_z$ is zero.  The $(I_z)^+$ and the
$(-I_z)^+$ states in these two sets can be grouped together.  In this
new grouping, the combined set contains 128 doubly-degenerate states
of the set
$\{$8$^+$,4(7$^+$),8(6$^+$),12(5$^+)$,16(4$^+$),20(3$^+$),25(2$^+$),27(1$^+$),\break
15(0$^+$)$\}$ at the excitation energy $E_x$=2$\Delta
E_{12}$=6$\epsilon_0$.

We have thus obtained the spectrum of the lowest-lying multiplets of
states. The spectrum of the higher states can be obtained in a similar
way. We can summarize the theoretical energy spectrum of the toroidal
$^{12}$C nucleus in Table \ref{tb6} and in Fig. \ref{fig2}.

\begin{table}[H]
\caption {The theoretical spectrum of $^{12}$C in a toroidal
  configuration.  Here, $E_x$=$E_I-E_0$, $E_I$ is the energy of a
  state in the multiplet, $E_0$ is the toroidal ground state energy,
  $\epsilon_0$=1.25 MeV by matching with the experimental spectrum,
  and $N$ is the number of doubly-degenerate states in the multiplet.
} \centering
\begin{tabular}{|c|c|l|l|c|}
\hline
 & & & & \\
Toroidal Ground State &  $\frac{E_x}{\epsilon_0}$  &$~~E_I$& ~~~~~~$I_z$=$I$ states & $N$\\
 \& Toroidal Multiplets&  &($\!$MeV$\!$)$\!$&   & \\
 \hline
 $[$(0p0h)$]^+$   & 0  & 7.654&  0$^+$    & 1 \\
$[$(1p1h)$_{12}]^-$ &   3    & 11.41&   0$^-\!\!$, 2($1^-\!)$,2(2$^-\!)$,2(3$^-\!$),$\!4\!^-$ & 8\\
$[$(1p1h)$_{02}]^+$ &   4    & 12.66&   1$^+\!\!$, 2($2^+\!$), 3$^+$  & 4   \\
$[$(2p2h)$_{12}]^+$ &   6    & 15.16&   3$(0^+\!$),5(1$^+\!$), 4(2$^+\!$),   & 18\\
& & & 2($3^+\!$),2(4$^+$),5$^+\!$,6$^+\!$  &\\
$[$(1p1h)$_{12\nu}$(1p1h)$_{12p}]^+$ &   6    & 15.16&15($0^+\!),$27$(1^+\!)$,25($2^+\!)$, & 128 \\
 & & &20(3$^+\!)$,16(4$^+\!)$,12($5^+\!)$,&\\
& & & 8(6$^+\!)$,4(7$^+\!)$,8$^+\!$&\\
$[$(1p1h)$_{12}$(1p1h)$_{02}]^-$ &   7    & 16.42& 4(0$^-\!)$,7(1$^-\!)$,5(2$^-\!)$,  &24\\
& & &3(3$^-\!)$,2($4^-\!)$,2($5^-\!)$,6$^-\!$& \\
$[$(1p1h)$_{12\nu}$(1p1h)$_{02p}]^-$ &   7    & 16.42&8(0$^-\!)$,15(1$^-\!$),12(2$^-\!)$,&64\\
& & & 9(3$^-\!$),8(4$^-\!$),7(5$^-$$\!\!$),&\\
& & &4(6$^-$$\!\!$),7$^-$&\\
$[$(1p1h)$_{13}]^+$ &   8    & 17.67&   1$^+\!\!$,2(2$^+\!$),2($3^+\!$),2(4$^+\!$),5$^+$&8 \\
$[$(2p2h)$_{02}]^+$ &   8    & 17.67&   0$^+\!$,1$^+\!\!$, 4$^+$ &3\\
$[$(1p1h)$_{02\nu}$(1p1h)$_{02p}]^+$ &   8    & 17.67&   6(0$^+\!)$,8(1$^+\!$),3(2$^+\!$),&32\\
& & & 4(3$^+\!$),6(4$^+\!$),4(5$^+\!$),6$^+\!$ &   \\
$[$(1p1h)$_{03}]^-$ &   9    & 18.92&   2$^-\!\!$, 2($3^-\!$),4$^-$&8   \\
$[$(3p3h)$_{12}]^-$ &   9    & 18.92&   0$^-\!\!$, 2(1$^-\!$),2(2$^-\!$),2(3$^-\!)$$\!$,4$\!^-$&8    \\
$[$(2p2h)$_{12\nu}$(1p1h)$_{12p}]^-$ &   9    & 18.92&    30($0^-\!\!$),59($1^-\!\!$),55($2^-\!\!$),  & 288 \\
  &  &  &  47($3^-\!\!$),36($4^-\!\!$), 25(5$^-\!\!$), &  \\
 & & & 16(6$^-\!\!$),10($7^-\!\!$),6(8$^-\!\!$), & \\
& & & 3(9$^-\!\!$),10$^-$ &   \\
$[$(2p2h)$_{12p}$(1p1h)$_{12\nu}]^-$ &   9    & 18.92&   
(same as above)  & 288 \\
$[$(2p2h)$_{12}$(1p1h)$_{02}]^+$ &   10    & 20.15 &   3(0$^+\!\!$),6(1$^+\!\!$),6(2$^+\!\!$),
& 24 \\
& & & 5(3$^+\!\!$),3($4^+\!\!$),5$^+\!\!$ & \\
$[$(2p2h)$_{12p}$(1p1h)$_{02\nu}\!]^+\!\!$ &   10    & 20.15 &  15($0^+\!\!$),30($1^+\!\!$),
29(2$^+\!\!$),
& 144 \\
& & &25(3$^+\!\!$),18(4$^+\!\!$),11(5$^+\!\!$), & \\
& & & 7(6$^+\!\!$),5(7$^+\!\!$),3(8$^+\!\!$),9$^+$  & \\
$[$(2p2h)$_{12\nu}$(1p1h)$_{02p}\!]^+\!\!$ &   10    & 20.15 &  (same as above)
& 144 \\
$[$(2p2h)$_{02}$(1p1h)$_{12}]^-$ &   11    & 21.40 &   4$^-\!\!$,2(3$^-\!\!)$,2(2$^-\!\!)$,2(1$^-\!\!)$,0$^-\!\!$
& 8 \\
$[$(4p4h)$_{12}]^+$ & 12    & 22.65 &   0$^+$   &1  \\
$[$(2p2h)$_{12\nu}$(2p2h)$_{12p}]^+$        & 12    &   22.65 & 69(0$^+\!\!$),130(1$^+\!\!$),117(2$^+\!\!$),      & 648   \\
       &  &   &  96(3$^+\!\!$),78(4$^+\!\!$),58(5$^+\!\!$),   &  \\ 
       &  &  &  42(6$^+\!\!$),26(7$^+\!\!$),16(8$^+\!\!$), &  \\ 
     &     &   & 8(9$^+\!\!$),5(10$^+\!\!$),2(11$^+\!\!$),12$^+$&  \\
$[$(3p3h)$_{12}$(1p1h)$_{02}]^-$ &   13    & 23.90 &   0$^-\!\!$,2(1$^-\!\!)$,2$^-$
& 4 \\
$[$(3p3h)$_{12\nu}$(1p1h)$_{02p}]^-$ &   13    & 23.90 &   8$(\!0^-\!\!)$,15$(\!1\!^-\!\!)$,12$(2\!^-\!\!)$,9$(3^-\!\!)$, & 64 \\
& & &  8$(4^-\!\!)$,7$(5^-\!\!)$,4$(6^-\!\!)$,7$^-$ & \\
$[$(3p3h)$_{12p}$(1p1h)$_{02\nu}]^-$ &   13    & 23.90 &   
(same as above)
& 64 \\
$[$(2p2h)$_{12}$(2p2h)$_{02}]^+$ & 14    & 25.15 & $0^+$,$1^+$,$2^+$&  3   \\
$[$(2p2h)$_{12\nu}$(2p2h)$_{02p}]^+$ & 14    & 25.15 & 13(0$^+\!\!$),23(1$^+\!\!$),20(2$^+\!\!$)    &  108   \\
       &  &  &  15(3$^+\!\!$),13(4$^+\!\!$),10(5$^+\!\!$),& \\
      &  &  & 7(6$^+\!\!$),3(7$^+\!\!$),2(8$^+\!\!$),9$^+$$\!\!$,10$^+$& \\
\hline
\end{tabular}
\label{tb6}
\end{table}

For simplicity, the single-particle energies in Eq.\ (\ref{eq3}) have
been simplified to depend only on $\Lambda_z^2/R^2$.  Such a
representation may be adequate for the lowest few toroidal shells.
However, in a more realistic case, the single-particle potential
should follow the toroidal density and should be a diffused potential
with a finite depth.  The higher toroidal shells are expected to be
unbound.  As a consequence, there will be a termination of the (npnh)
toroidal excitations, indicated by the absence of bound toroidal
particle-hole excitations at high energies and a rapid decrease of the
density of toroidal particle-hole excitation states.  It will be of
interest to search for the excitation energy maximum in $^{12}$C at
which the toroidal particle lie in the continuum and there will be no more 
bound toroidal particle-hole 
excitations.

\section{Comparison of  Toroidal  Signature 
 with  Experimental $^{12}$C Spectrum }

In the last few sections, we show that the $^{12}$C nucleus in a
toroidal configuration possesses a distinct spectrum of with well
defined spins, parities, and excitation energies.  They arise from
particle-hole excitations from a toroidal $\Lambda_i$-shell to another
toroidal $\Lambda_f$-shell and represent the signature of the $^{12}$C
nucleus in the toroidal configuration.  Using such a signature, we
explore whether there may be toroidal states in $^{12}$C build on the
$0^+$ Hoyle state at 7.654 MeV as the head of the toroidal band.

In our exploratory survey, we expect that the toroidal states of $^{12}$C and the
low-lying toroidal multiplets will show up as $^{12}$C resonances among states that have
a large probability to breakup into 3 alpha particles.  
We envisage that
for a system that breaks into three alpha particles, 
an emitted alpha particle will be in contact with a Be nucleus
at the moment of scission.   The Be nucleus will likely be in the form of a two-alpha cluster.
 A rod configuration with a three-alpha cluster lies at around 16.5 MeV \cite{Ren18}. 
The configuration of the $^{12}$C system that decays into three alpha particles at
excitation energies below the onset of the rod configuration at about 16.5 MeV  
 is likely to consist of a triangular cluster of   alpha particles in various degrees of contact.    
Such a triangular clusters of three alpha particles has a probability
amplitude overlap with the toroidal configuration and 
can originate from  the evolution of 
a  toroidal configuration under sausage deformation of order $\lambda$=3.
Accordingly, we search for good candidate toroidal states in
$^{10}$B($^{3}$He,p)$^{12}$C$^*$$\to$3$\alpha$ and
$^{11}$B($^{3}$He,d)$^{12}$C$^*$$\to$3$\alpha$ reactions
\cite{Kir10,Kir12,Alc12} as shown in Figs.\ \ref{fig3}, \ref{fig4},
and \ref{fig5}.  We shall discuss the results of our search in the
following subsections.

\subsection{ Comparison of   $^{11}$B($^3$He,d)$^{12}$C$^*$$ \to$$ 3\alpha$ Spectrum with 
 Toroidal (1p1h)$_{12}^-$  and (1p1h)$_{02}^+$ Multiplets   }

\begin{figure} [H]
\hspace*{0.3cm}
\includegraphics[scale=0.50]{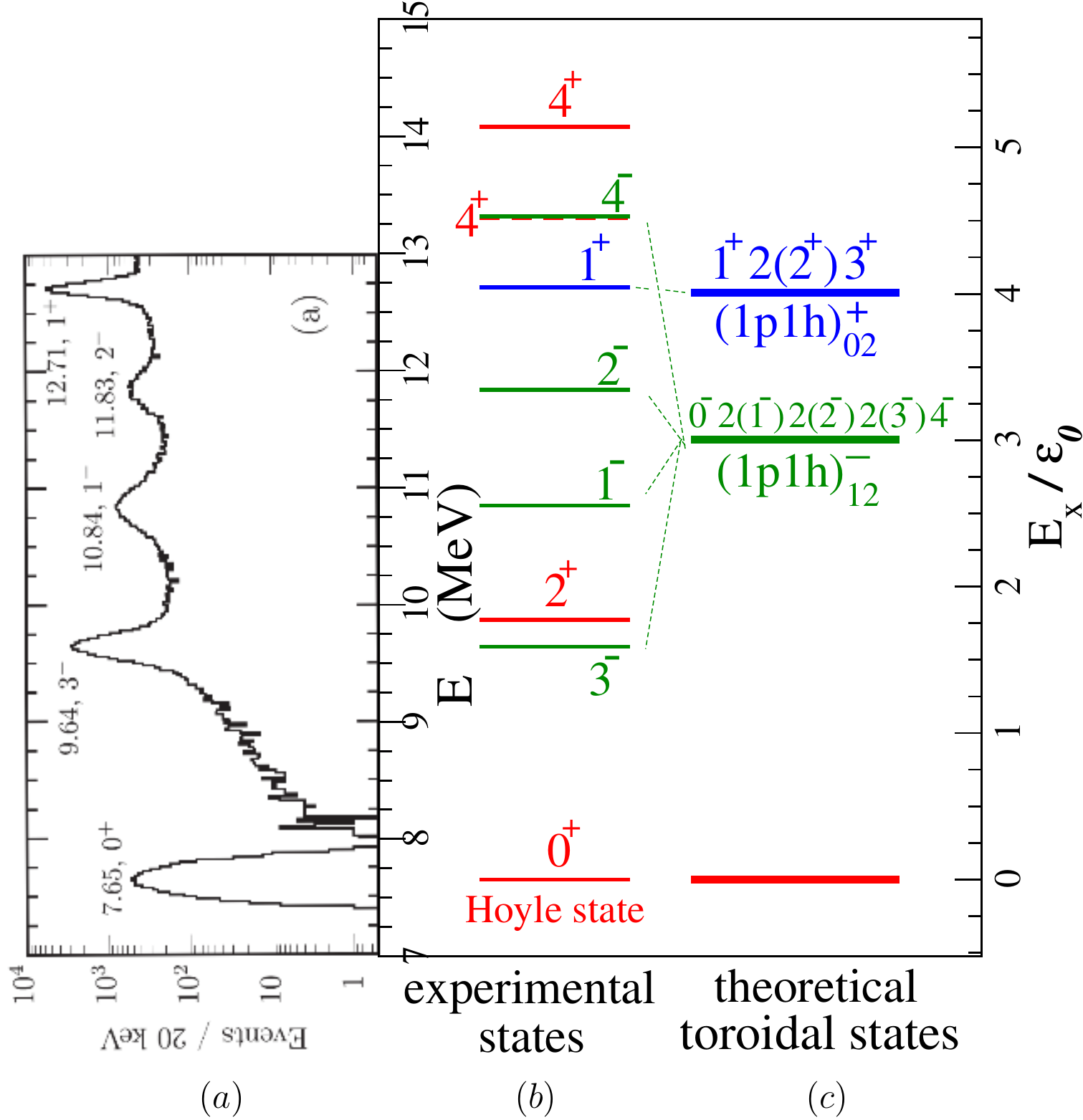} 
\caption{(color online).  (a) Experimental excitation function for the
  reaction $^{11}$B($^{3}$He,d)$^{12}$C$^* $$\to$ 3$\alpha$ in
  logarithmic scale as a function of the $^{12}$C$^*$ excitation
  energy $E$ in the range of $7 \le E\le$ 13 MeV obtained by Kirsebom
  $et~al.$ \cite{Kir10}, with the axis of the excitation energy on the
  left.  (b) Experimental level scheme from the compilation of
  \cite{Kel17}.  (c) Theoretical level scheme of toroidal states with
  the axis of $E_x=(E-E_0)/\epsilon_0$ on the right.  Subsequent
  matching between theoretical and experimental excitation energies
  leads to $\epsilon_0$=1.25 MeV.  }
\label{fig3}
\end{figure}

\begin{figure} [H]
\centering
\hspace*{-0.3cm}
\includegraphics[scale=0.60]{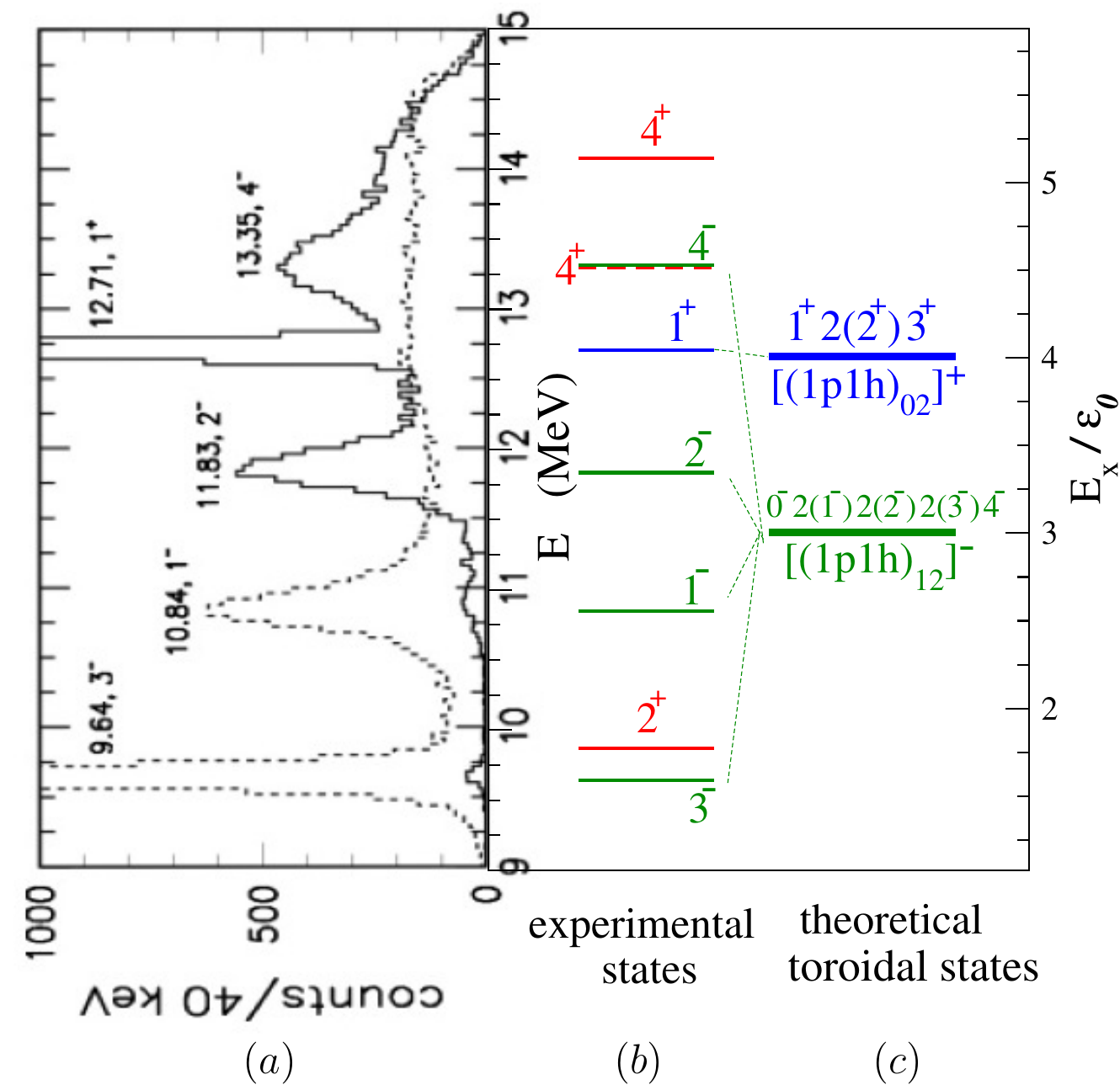} 
\caption{(color online).  Experimental excitation function for the
  reaction $^{11}$B($^{3}$He,d)$^{12}$C$^* $$\to$ 3$\alpha$ in linear
  scale as a function of the $^{12}$C$^*$ excitation energy $E$ in the
  range of $9 \le E\le$ 15 MeV obtained by Kirsebom $et~al.$
  \cite{Kir10}, with the axis of the excitation energy on the left.
  (b) Experimental level scheme from the compilation of \cite{Kel17}.
  (c) Theoretical toroidal states with the axis of
  $E_x/\epsilon_0$=$(E-E_0)/\epsilon_0$ on the right.}
\label{fig4}
\end{figure}

In the experiments of Kirsebom $et~al.$ \cite{Kir10,Kir12}, excited
intermediate $^{12}$C$^*$ states were produced by the bombardment of
the $^{3}$He projectile at 8.5 MeV onto a $^{11}$B target by the
$^{3}$He + $^{11}$$\text{B } \to d + {}^{12}$$\text{C}^*$ reaction,
with the subsequent breakup of the $^{12}$C$^*$ intermediate state
into three alpha particles, $^{12}{\text C}^* \to 3 \alpha$.  The
complete kinematic data of the final deuterium $d$ and the 3 alpha
particles have been recorded with detectors of fine resolutions and
segmentation to allow the determination of (i) the energy of the
intermediate $^{12}$C$^*$ state, (ii) the history of its subsequent
decay, and (iii) the energy distribution Dalitz plot of the 3 alpha
particles.  From these pieces of information, the spins and parities
of the prominent $^{12}$C resonances can be inferred.  The complete
kinematic data allow the removal of most of the random coincidences
and decay channels that do not involve the production of intermediate
excited states of $^{12}$C$^*$.  The measurement is essentially free
of background \cite{Alc09}.  The spectrum as shown in
Figs.\ \ref{fig3}(a) and \ref{fig4}(a) from Kirsebom $et~al.$
\cite{Kir10,Kir12} can be considered to be the spectrum for the
produced intermediate excited $^{12}$C$^*$ states, which include both
the identified $^{12}$C resonances as well as unresolved and
un-identified excited $^{12}$C$^*$ states, as we shall discuss below.

It is instructive to review how (i) the $^{11}$B($^3$He,d)$^{12}$C$^*$
reaction mechanism, (ii) the selection of the $^{12}$C$^*\to 3\alpha$ final states, and
(iii) the knowledge of the toroidal nucleus structure information help
guide us in the search for toroidal $^{12}$C particle-hole excitation
states.  The ($^3$He,d) process of \cite{Kir10,Kir12} strips a proton
from the incident projectile nucleus, $^{3}$He, turns it into a
deuterium, and deposits the stripped proton onto the target $^{11}$B
nucleus, which has 5 protons and and 6 neutrons.  With this addition
of the stripped proton, the system with 6 protons and 6 neutrons
completes a doubly-closed shell for a toroidal shape.  
By the selection of the breakup into three
alpha particles, we judiciously search for the energy location for the $^{12}$C
toroidal state for processes in which the stripped proton can excite
and polarize the nucleons in the $^{11}$B system to re-configure
themselves into a doubly closed shell toroidal nucleus with only a
single unoccupied state in the proton $\Lambda$=1 shell, and the
stripped proton can then fill up the unoccupied proton single-particle
state to lead to the (0p0h) toroidal $^{12}$C ground state.

With the selection of breaking up  into 3 alpha particles, we can look for
events at another higher energy for the toroidal (1p1h)$_{12}^-$ $^{12}$C
state, in which the stripped proton excites the nucleons in the $^{11}$B
system to re-configure themselves into a toroidal configuration with a
hole in the $\Lambda$=1 shell, while the stripped proton goes on to an
unoccupied proton state in the higher unoccupied $\Lambda$=2 shell, leading to the
(1p1h)$_{12}^-$ state of toroidal $^{12}$C.  We expect that at the
appropriate energies, the
$^{11}$B($^{3}$He,d)$^{12}$C$^*$$\to$3$\alpha$ reaction should
favorably populate (0p0h) and (1p1h)$_{12}^-$ states of the toroidal
$^{12}$C nucleus.  In a similar manner, the (1p1h)$_{02}^+$ multiplets
of $\{1^+,2(2^+),3^+\}$ in Table \ref{tb6} should be likewise
favorably excited in the
$^{11}$B($^{3}$He,d)$^{12}$C$^*$$\to$3$\alpha$ reaction.

Figure \ref{fig3}(a) gives the experimental excitation function for
the $^{11}$B($^{3}$He,d)$^{12}$C$^*$$\to$3$\alpha$ reaction in the
range $7\le E\le 13$ MeV in logarithmic scale, and Fig.\ \ref{fig4}(a)
shows the same excitation function in the range $9\le E\le 15$ MeV in
linear scale.  Figs.\ \ref{fig3}(b) and \ref {fig4}(b) show the energy
levels from the experimentally identified $^{12}$C excited states
from the compilation of \cite{Kel17}.  The excitation energy $E$
relative to the energy of $^{12}$C ground state is given on the left
axis.

The experimental excitation functions in Figs.\ \ref{fig3}(a) and
\ref{fig4}(a) indicate that the $0^+$ Hoyle state at $E$=7.654 MeV is
prominently excited, as are the $3^-$ state at 9.654 MeV, the $1^-$
state at 10.847 MeV, the $2^-$ state at 11.837 MeV, the $4^-$ state at
13.314 MeV, and the $1^+$ state at 12.71 MeV.  By comparing the
experimental and the theoretical spectrum in Figs.\ \ref{fig3}(c) and
\ref{fig4}(c), we find that the lowest theoretical multiplet in
toroidal $^{12}$C contains states with spins and parities that
coincide with those of the experimental states.  It is therefore
reasonable to identify the $0^+$ Hoyle state at 7.654 MeV to be the
(0p0h) ground state of the toroidal configuration and the set of
lowest $\{3^-$(9.654 MeV), 1$^-$(10.847 MeV), 2$^-$(11.837 MeV),
4$^-$(13.314 MeV)$\}$ states to be members of the (1p1h)$_{12}^-$
multiplet.  Following such an identification, we set the average
excitation energy of the four states $\{3^-, 1^-, 2^-, 4^-\}$ at 11.41
MeV to be the excitation energy of the (1p1h)$)_{12}^-$ multiplet.
Such a matching leads to the theoretical energy scale,
\begin{eqnarray}
\epsilon_0= 1.25 ~~{\rm  MeV }.
\end{eqnarray}
By the definition of $\epsilon_0$ as $\hbar^2/2mR^2$, we obtain the
corresponding effective major radius,
\begin{eqnarray}
R=4.06 ~ {\rm fm}.
\end{eqnarray}
The root-mean-squared radius $r_{\rm rms}$ of
the Hoyle state as determined in previous 3$\alpha$ cluster
theoretical models: $r_{\rm rms}$=3.56 fm from the resonating group
method in Ref.\ \cite{Kam81}, $r_{\rm rms}$=3.40 fm from the generator
coordinate method in Ref.\ \cite{Ueg79}, $r_{\rm rms}$=3.83 fm from the
generator coordinate method in Ref.\ \cite{Fun05}, and $r_{\rm
  rms}$=3.82 fm from the three-body quantum dynamics in
Ref.\ \cite{AR07}.  The
value of $R$ extracted from the (one-particle
one-hole) spectrum here is slightly higher than 
the theoretical $r_{\rm rms}$ determined by cluster models
calculations for the Hoyle state.   Such a difference may arise partially from the fact that  
the particle-hole excitation involves the excitation to the unoccupied $\Lambda=2$ single-particle orbital
which has naturally a  greater radius compared to the $\Lambda=0$ and $\Lambda=1$ occupied orbitals in the Hoyle state. 

The knowledge of $\epsilon_0$ allows the determination of the
theoretical toroidal spectrum of $^{12}$C as tabulated in Table
\ref{tb6} and shown in Figs.\ \ref{fig2}-\ref{fig5}.  In particular,
we find that the $1^+$ state at $E_z$=12.71 MeV approximately matches
the $1^+$ state in the multiplet of (1p1h)$_{02}^+$.

We note in Figs.\ \ref{fig3} and \ref{fig4} that  there are
also a $2^+$ state at 9.87 MeV, a $4^+$ state at 13.3 MeV \cite{Fre11} and
another $4^+$ state at 14.079 MeV \cite{Kel17}.   
These states  are collective
rotational states that are not excited significantly by the stripping
$^{11}$B($^{3}$He,d)$^{12}$C$^*$$\to$3$\alpha$ reaction.
The
analysis of the interacting alpha particle  model  \cite{Gar15}  and the algebraic cluster model \cite{Bij00,Bij02}  found that the sequence of
the 0$_1^+$(ground state), 2$_1^+$(4.33 MeV), and 4$_2^+$(14.079 MeV)
form the rotational states of the ground band, whereas the sequence of
the Hoyle state $0_2^+$(7.654 MeV), $2_2^+$(9.870 MeV), and
$(4_1^+$)(13.3 MeV) state \cite{Fre11} form a rotational band with a
different moment of inertia.  With the identification of these
rotational bands built on the  Hoyle state and the ground state,
a  consistent description  of the properties  of all lowest eleven identified states of the $^{12}$C nucleus emerges. Namely,
in addition to the rotational $0^+,2^+,4^+$ bands of the ground state and the Hoyle state, the other low-lying states  can be attributed simply  to one-particle one-hole toroidal multiplet excitations from a spatially extended Hoyle state in the toroidal configuration.

\subsection{ Excitation Strength 
within the (1p1h)$_{12}^-$ Multiplet}

The excitation function of the stripping reaction $^{11}$B($^{3}$He,d)$^{12}$C$^*$
leading to toroidal (1p1h)$_{12}^-$ particle-hole states in
$^{12}$C$^*$ depends on the degree of occupation of  the $\Lambda$=1 shell and the degree of emptiness of the
unoccupied $\Lambda$=2 shell, which are the same for all members of
the shell-to-shell particle-hole multiplet.  Consequently, all members
of the (1p1h)$_{12}^-$ multiplet should be approximately equally
produced.  An examination of the widths and the heights of the identified
lowest $\{ 1^-, 2^-, 4^-\}$ states above the underlying unresolved
states indicate that these three are approximately equally populated,
supporting their identification as members of the (1p1h)$_{12}^-$
multiplet.

We note in passing that the experimental strength for the excitation
of the (0p0h) Hoyle state in Figure \ref{fig3} is smaller than the
strength for the excitation of the (1p1h) states in the
$^{11}$B($^{3}$He,d)$^{12}$C$^*$ reaction.  In stripping a proton from
the incident $^{3}$He nucleus and depositing the proton onto the
$^{11}$B nucleus, the reaction of $^{11}$B($^{3}$He,d)$^{12}$C$^*$ is
peripheral in nature.  The dominant contribution to the reduced DWBA
cross section depends on the tail of the radial single-particle
bound-state wave functions of the deposited single-particle state
inside the nucleus.  The (0p0h) excitation deposits a proton to the
deeper $\Lambda=1$ shell single-particle state whereas the
(1p1h)$_{12}^-$ excitation deposit the proton to the higher
$\Lambda=2$ shell single-particle states with a greater magnitude of
the wave function in the peripheral region of the nucleus.  Therefore,
the excitation strengths for the (1p1h)$_{12}^-$ (3$^-$, 1$^-$, 2$^-$,
4$^-$) multiplet is higher than that for the (0p0h) $0+$ Hoyle state.

\subsection{ Number of  Toroidal States in Multiplets and the Presence of Underlying Broad Structure}

Our comparison with identified resonances reveals that there appear to
be many more theoretical states than the number of experimentally
identified states.  On the other hand, the excitation function in
Fig.\ \ref{fig3}(a) in
the  measurement of \cite{Kir10}
possesses strengths away from the cleanly
identified peaks.  The  measurement of \cite{Kir10} 
is 
free of background \cite{Alc09}
because the complete kinematics 
satisfying the conservation of energy and momentum
allow the identification all participating initial and final particles, and   noise production of the 3$\alpha$+$d$ assembly will not satisfy the energy and momentum conservation.  
Therefore,  the presence of these extra
excitation strength in Fig.\ \ref{fig3}(a)
obtained in   \cite{Kir10} 
 indicates that additional
overlapping resonances of $^{12}$C have been produced in the
$^{11}$B($^3$He,d)$^{12}$C$^*$$\to$3$\alpha$
measurement.  They constitute non-vanishing strengths in the Dalitz
plot and represent broad and unidentified excited states of $^{12}$C.
They are likely the remaining members of the toroidal multiplets that
are also produced.  If they indeed are, then their number should match the
remaining number of unidentified states.  We can make a rough estimate
of the remaining number of unidentified states as follows.  Each state
of the multiplet is expected to be produced in approximately an equal
single-particle ``particle-hole" strength in the stripping
$^{11}$B($^3$He,d)$^{12}$C$^*$$\to$3$\alpha$ reaction.  The excitation
function strengths of the resolved resonances in the range can be used
as a yard stick to estimate the number of the remaining produced
un-identified $^{12}$C$^*$ resonances in the underlying broad
structure in Fig.\ \ref{fig4}(a).
 
In the data of Fig.\ \ref{fig4}(a), the $1^-$(10.847 MeV),
$2^-$(11.837 MeV), and $4^-$(13.314 MeV) resonances are members of the
(1p1h)$_{12}^-$ multiplet arising from single-particle excitations
populating a one-particle-one-hole state in the $\Lambda$=2 shell.

The strength of each the 
 $1^-$(10.847 MeV),
$2^-$(11.837 MeV), and $4^-$(13.314 MeV) resonances 
as suggested members of the
(1p1h)$_{12}^-$ multiple
defines roughly a single-particle ``particle-hole" unit, with
approximately the same area for each of the three peaks, and each
particle-hole unit leads to the production of one $^{12}$C$^*$
particle-hole state.  On that basis of using that (average) area as a
yardstick, we find that in the energy range of Fig.\ {\ref{fig4}(a) up
  to the instrumental cut-off excitation energy of $\sim$ 14.5 MeV,
  there are approximately 3 to 5 units of particle-hole strength for
  natural parity states and 2 to 4 units for the un-natural parity
  states in the underlying overlapping resonances, with a considerable
  degrees of uncertainty in these numbers due to the uncertainty in
  separating out the resolved peaks from the underlying broad
  structure.  The numbers of remaining
  un-identified members in the multiplet fall within the range of
  numbers of particle-hole states estimated to be present in the
  underlying broad structure in Fig.\ \ref{fig4}(a).  By this
  comparison, it appears that the total number of identified and
  unidentified resonances produced in the
  $^{11}$B($^3$He,d)$^{12}$C$^*$$\to$3$\alpha$ reaction matches
  approximately the total number of particle-hole states of the
  (1p1h)$_{12}^-$ and (1p1h)$_{02}^+$ multiplets.

\subsection{The Excitation Function in  $^{10}$B($^3$He,p)$^{12}$C$^*$$\to$$3\alpha$  
and  the Density of  Toroidal States at High Energies}

In another experiment, Alcorta $et~al.$\cite{Alc12} used the $^{3}$He
projectile to collide with a $^{10}$B target nucleus to study the
excited states of $^{12}$C by the
$^{10}$B($^3$He,p)$^{12}$C$\to$3$\alpha$ reaction at a beam energy of
4.9 MeV.  The complete kinematics of all four final particles were
collected using detectors of fine resolution and segmentation.  Again,
the full knowledge of the complete kinematics facilitate the
assignment of spins and parities.  Contributions from direct reactions
leading to the production of intermediate states other than
$^{12}$C$^*$ have been eliminated as much as possible so that the
excitation function of the 3$\alpha$ spectrum can also be considered
to be essentially free of background, pending future removal of the
small
$^{3}$He+$^{10}$B$\to$$^{8}$Be+$^{5}$Li$\to$$^{8}$Be+$p$+$\alpha$
and
$^{3}$He+$^{10}$B$\to$$\alpha$+$^{9}$B$\to$$\alpha$+$\alpha$+$^{5}$Li
contributions \cite{Alc09}.
 
\begin{figure} [h]
\hspace*{-0.4cm}
\includegraphics[scale=0.33]{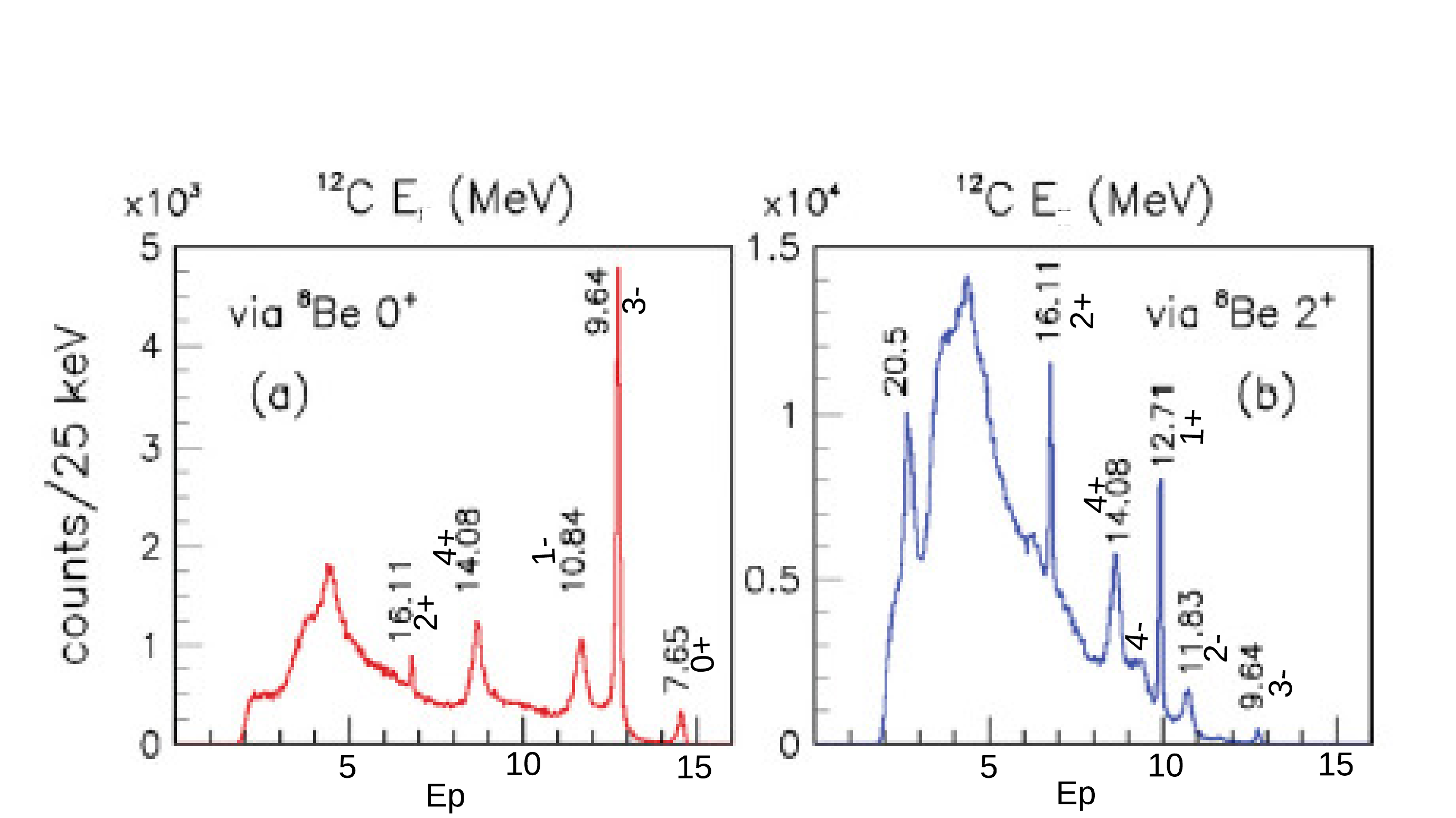} 
\caption{(color online).  Experimental excitation functions of the
  reaction $^{10}$B($^3$He,p)$^{12}$C$^*$$\to$3$\alpha$ as a function
  of the proton energy $E_p$ from Alcorta $et~al. $\cite{Alc12}.  The
  excitation energies $E_x$, spins, and parities of identified
  resonances of the excited $^{12}$C$^*$ states are indicated.
 Please note the different
  count scales on the $y$-axes in Figs. (a) and (b).  The decay of
  the excited $^{12}$C$^*$ state goes through the $0^+$ state of $^8$Be in (a) and
  through the $2^+$ state of $^8$Be in (b).  }
\label{fig5}
\end{figure}
The ($^3$He,p) reaction strips a neutron and a proton from the $^3$He
projectile and deposits the two nucleons onto the $^{10}$B target. It
is expected to populate favorably the (0p0h), (1p1h), and (2p2h)
toroidal excitations of $^{12}$C$^*$ at different excitation energies.
Experiment data of the $^{10}$B($^3$He,p)$^{12}$C$\to$3$\alpha$
reaction \cite{Alc12} in Fig.\ \ref{fig5} in the three alpha breakup
indicate that the Hoyle state, the lowest $\{1^-,2^-,3^-,4^-\}$ and
$1^+$ states are prominently produced.

  The additional possibility of the (2p2h) excitations with the
  $^{10}$B($^3$He,p)$^{12}$C$^*$$\to$$3\alpha$ reaction leads to an
  enhanced excitation function at higher excitation energies.  It is
  interesting to note that the experimental 4$^+$(14.08 MeV) state and
  the $2^+$(16.11 MeV) falls in the vicinity of the theoretical
  (2p2h)$_{12}^+$ multiplet whose members include a $2^+$ and a $4^+$
  state.  Whether these two states can be identified as two members of
  the (2p2h)$_{12}^+$ multiplet will require further theoretical and
  experimental investigations.  In Fig. \ref{fig5} the produced
  resolved states appear as sharp peaks on top of an underlying broad
  structure.  If we consider the sharp peaks as arising from a 2p2h excitation and we
employ the earlier method using the areas of the
  excitation function covered by known resonances as a
  ``two-particle-two-hole" (2p2h) yard stick to estimate the number of
  (2p2h) states involved in the underlying broad structure, we would
  come up with the result that the number of similar (2p2h) states
  comprising the underlying broad structure in Fig.\ \ref{fig5} is an
  order of magnitude greater than number of resolved and identified
  resonances.  Thus in addition to the few resolved and
  identified states that have been mentioned as possible members of
  the toroidal (2p2h) multiplets, the underlying structure represents
  also 
a large number of 
possible produced members of the toroidal multiplets in the
  $^{10}$B($^3$He,p)$^{12}$C$^*$$\to$3$\alpha$ reaction of
  Ref.\ \cite{Alc12}.

\begin{figure} [h]
\centering
\includegraphics[scale=0.33]{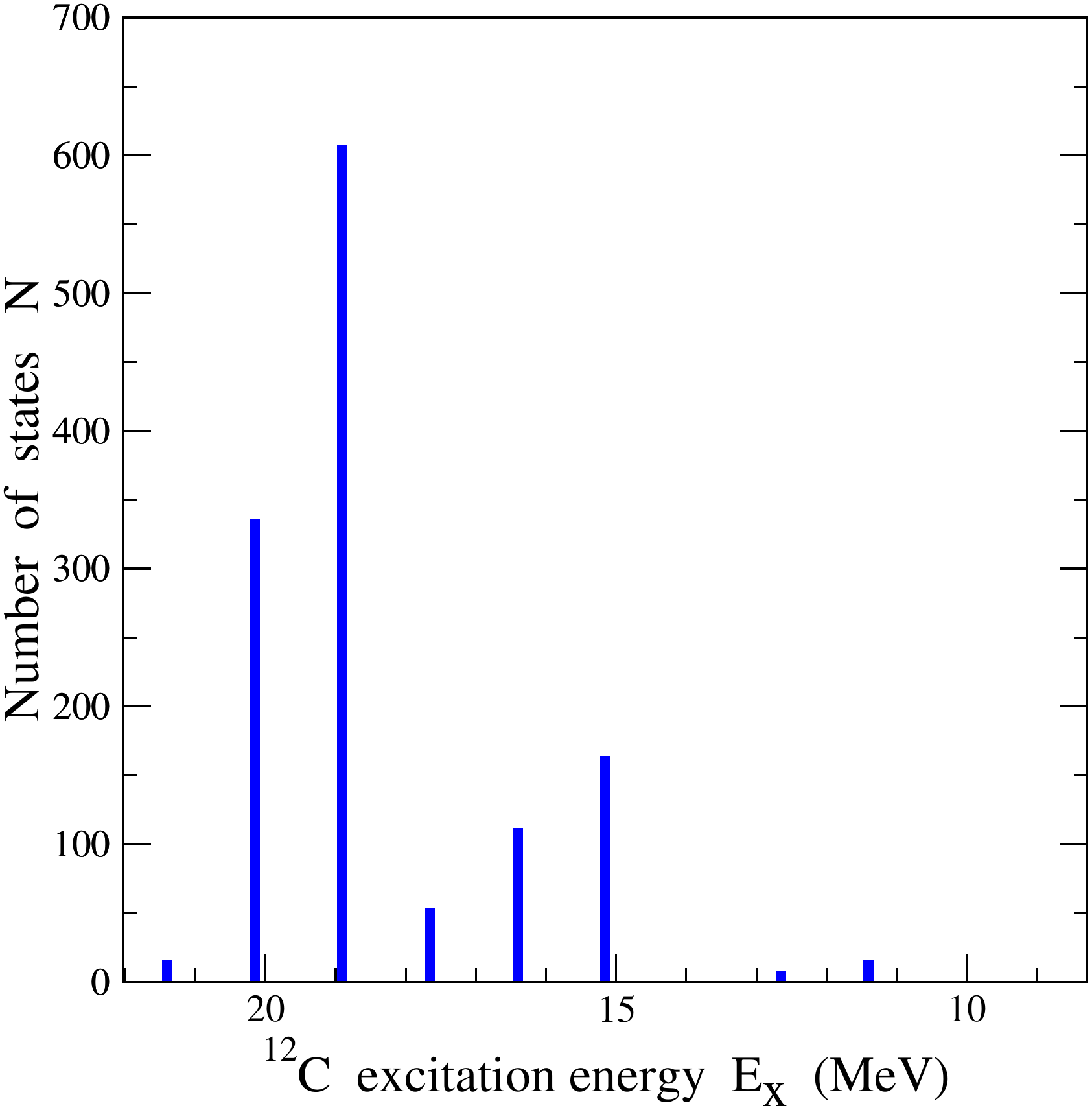} 
\caption{(color online).   Number of toroidal states as a function of the excitation energy  $E_x$ relative to the ground state for the $^{12}$C nucleus in a toroidal configuration.
}
\label{fig6}
\end{figure}
 
The experimental excitation function in Fig.\ \ref{fig5} provides
valuable information on the number of excited
$^{12}$C$^*$ states as a function of $E_x$.  
It indicates that the number  of $^{12}$C $^*$ states at $E_x$$\sim$ 10-20 MeV that decay 
into 3$\alpha$  is very large.   
Above the Hoyle state energy, the  toroidal particle-hole excitations 
built on the Hoyle state provide 
the dominant contribution to the number of states, on account of the
extended geometrical size of the Hoyle state.    It is illuminating to compare
the number of toroidal particle-hole states as a function of $E_x$  with
the general behavior of the  excitation function
for $^{12}$C$^*$ production in Fig.\ \ref{fig5}.
To facilitate such a comparison, we have
plotted the number of toroidal states as a function of 
$^{12}$C  excitation energy $E_x$  in Fig.\ \ref{fig6} along the direction of decreasing
$E_x$.  The number of toroidal
states is quite large at $E_x$$\sim$ 18-20 MeV and has a smaller peak
at $\sim$15-16 MeV.  Experimental excitation function in Fig.\
\ref{fig5} has a large peak at about 18 MeV and a substantial yield at
around 16 MeV.  There appears to be a qualitative correlation between
the shape of the experimental excitation function with the theoretical
variation of the number of toroidal states, but more work is needed to 
make a quantitative comparison.

From the above comparisons outlined in subsections A, B, C, and D,  there appear to be an approximate agreement of many pieces of data with 
the signatures of the $^{12}$C nucleus in the toroidal configuration, but there are
also many items that need to be further investigated to confirm the presence of such a  toroidal configuration.  
Subject to these further experimental and 
theoretical  tests that are opened up by such the  new suggestion,
 the Hoyle state and many of its excited states
may be tentatively identified as states of the $^{12}$C nucleus in the
toroidal configuration.

\section{Toroidal ${}^{12}$C in Toroidal Constraint Dynamics}

Our investigation points to the need to devise tools within the
mean-field theory that can constrain the nucleus so to
possess local toroidal energy equilibrium configurations.   We
should be prepared to examine the toroidal configuration as excited
diabatic states of the system.  Success for formulating such tools
will be useful not only for the $^{12}$C nucleus but also for the many
diabatic states that may be associated with the large region of
toroidal high-spin isomers in $\alpha$-conjugate nuclei up to $A\le
70$ \cite{Sta15,Sta15a,Sta16}.  It will also be useful for the
investigation of diabatic toroidal states in the intermediate and
superheavy mass region for which some recent progress has been made
\cite{Won18,Kos17,Sta17}.

We consider a Hamiltonian $H_0$ which contains already many
constraints that have been imposed on the system.  We wish to impose
an additional toroidal constraint to hold the nucleus in a toroidal
shape.  For this purpose, we introduce a radial moment $\sigma_\rho^2$
to characterize the radial property of the density distribution that
can constrain the density into a toroidal shape,
\begin{eqnarray}
\sigma_\rho^2=\left \{ \int d{\bb r} ~n({\bb r})\left  (\rho- \langle \hat \rho\rangle \right  )^2\right \},
\label{eq19}
\end{eqnarray}
where $\langle \hat  \rho\rangle$ is the expectation value of the coordinate 
$\hat \rho$=$\rho$=$\sqrt{x^2+y^2}$,
\begin{eqnarray}
\langle \hat \rho\rangle=\int d{\bb r}  \sum_{i=1}^{A} \psi_i^*(\bb r)~\hat \rho ~ \psi_i(\bb r),
\end{eqnarray}
and $n(\rho,z)$ is the nuclear density which can be determined
self-consistently from the set of the wave functions $\{ \psi_i\}$ of
occupied states
\begin{eqnarray}
n(\bb r)=\sum_{i=1}^{A} \psi_i^*(\bb r) \psi_i(\bb r).
\end{eqnarray}
The constraint of a fixed toroidal radial moment $\sigma_\rho^2$,
 can be imposed with the  additional  Lagrange multiplier $\lambda$, 
\begin{eqnarray}
H=H_0 + \lambda\left \{ \int d{\bb r} ~n(\bb r)\left  (\rho- \langle \hat \rho\rangle \right  )^2-\sigma_\rho^2\right \}
\label{eq30}
\end{eqnarray}

Upon a minimization of the Hamiltonian $H$ with respect to a variation
in $\psi_i^*$, we have
  \begin{eqnarray}
\frac{\delta H}{\delta \psi_i^*}&&=\frac{\delta H_0}{\delta \psi_i^*}
+ \lambda \int d{\bb r} ~\left  (\rho- \langle \hat \rho\rangle \right  )^2 \psi_i
\nonumber\\
&&+({\rm terms ~involving ~~}\frac{\delta \langle \hat \rho \rangle}{\delta \psi_i^*})
\end{eqnarray}
To the extent that the change of the bulk $\langle \hat \rho\rangle$
with respect to the change of individual single-particle states
$\psi_i^*$ is small when the occupation numbers of the states have
settled down and do not change, the last term of the above variation
can be neglected.  Writing $H_0$ in terms of the single-particle
Hamiltonian $h_0$ as
\begin{eqnarray}
H_0=\int d{\bb r} ~ \sum_{i=1}^{A} \psi_i^*({\bb r})~h_0 ~\psi  ({\bb r}),
\end{eqnarray}
and we have
\begin{eqnarray}
\frac{\delta H_0}{\delta \psi_i^*} = \int d{\bb r} ~h_0 \psi_i .
\end{eqnarray}
Under the toroidal constraint of Eq.\ (\ref{eq30}),
the minimization of $H$ with respect to $\psi_i^*$ leads to  
the single-particle Hamiltonian  under the toroidal constraint 
\begin{eqnarray}
h'=h_0 + \lambda ~\left (\rho -  \langle \hat \rho\rangle \right  )^2.
\label{21}
\end{eqnarray}
In the case with a large $R/d$ ratio, $ \langle \hat \rho \rangle \sim
R$.  Thus, we observe that the last term of Eq.\ (\ref{21}) is
approximately in the same form as the toroidal potential of
Ref.\ \cite{Won73}, with the Lagrange multiplier $\lambda$ playing the
role of the harmonic oscillator frequency $\omega_\perp^2$ multiplied
by the nucleon mass $m$.  We obtain the result that with the addition
of the toroidal constraint Eq.\ (\ref{eq19}), the variation principle
lead to a single-particle toroidal potential that is similar to the
potential of \cite{Won72,Won73} with the $\omega_\perp$ appearing as a
variational parameter.  For a given quadruple moment that leads to the
proper $R$, the variation $\lambda$ will lead to the proper radial
width $d$ of the transverse degree of freedom.

Another approach to examine toroidal nuclei can be carried out in
diabatic mean-field calculations \cite{Zha10}.  This is achieved by
constraining the occupation of the single-particle states so that at
the locations of the crossing of two single-particle states at the top
of the fermi surface, one does not choose to occupy the state of the
lowest energy.  Instead, one maintains the diabatic configuration with
the occupation of the state with the highest overlap with the earlier
state before the level crossing, leading to a diabatic energy
equilibrium as in \cite{Zha10}.  Diabatic calculations have been
performed successfully for $^{24}$Mg \cite{Zha10} and $^{28}$Si
\cite{Cao19}.  

It will be of interest to see whether constraint dynamics or diabatic
calculations for $^{12}$C will reveal the toroidal $^{12}$C states as
discussed here.  Although such constraint dynamics and diabatic calculations may appear simple in theoretical  formulation, their implementation 
onto the self-consistent mean-field computer program are difficult tasks.  We shall examine the
microscopic mean-field description using the simpler method with variational wave functions.

\section{ Microscopic Description of $^{12}$C States using Variational Wave Functions}

The phenomenological analysis in Section V suggests the possibility
that the Hoyle state and many of its low-lying excited states may be
tentatively attributed to the particle-hole excitation of the $^{12}$C nucleus from a
toroidal configuration. It is therefore of great interest to search
for the microscopic foundation for such a toroidal description for the Hoyle state
as the band head of the particle-hole excitations,  starting with
the mean-field approximation using the Skyrme energy-density
functional \cite{Vau72,Vau73,Bar82}.  

Assuming a $^{12}$C nucleus with an intrinsic axial symmetry, we
describe the states of the nucleus by a Slater determinant of neutrons
and protons occupying the lowest-energy single-particle states
$|n_\rho$, $n_z$, $\Lambda_z$, $\Omega_z \rangle $.  For $^{12}$C, the
lowest-energy single-particles states are $|0,0,0,\pm 1/2\rangle$,
$|0,0,1,\pm 3/2\rangle$, and $|0,0,1,\pm 1/2\rangle$. 
 Limiting our considerations to $n_\rho$=0 and $n_z$=0
states, we shall omit the labels of ``$n_\rho$=0" and ``$n_z$=0"
for brevity of
notation.  The
variational parameters $(R,d_\rho,d_z)$ of the single-particle wave
functions will be so chosen that they allow the examination of the
energy surface of the nucleus in many different configurations : the
spherical, prolate spheroid, oblate spheroid, bi-concave disks, and
toroidal configurations.  For simplicity, we shall assume the same set
of spatial and spin wave functions for neutrons and protons and
neglect the spin-orbit interaction so that the radial wave function
${\cal R}_{\lambda_z \Omega_z}$ of the single-particle state
$|\Lambda_z \Omega_z\rangle $ depends only on $\Lambda$, the absolute
value of $\Lambda_z$.  The wave functions of the occupied
single-particle states for a proton or a neutron can be represented in
terms of the variation parameters $(R,d_\rho,d_z)$ by
\begin{subequations}
{
\begin{eqnarray}
&&\Psi_{\Lambda_z \Omega_z}(\rho,z,\phi)={\cal R}_{\Lambda}(\rho )Z(z) [\Phi_{\Lambda_z} (\phi)\chi_{s_z}]^{\Omega_z},
~~~~
\label{Psi}
 \\
\hspace{-0.7cm} \text{where~} &&~~ {\cal R}_{\Lambda} (\rho)=N_{\Lambda}\rho^{\Lambda}
\exp \left \{  -\frac{(\rho-R)^2}{2(d_\rho^2/\ln 2)}\right \},~~~~
\label{R}
\\
&&~~Z(z)=N_Z \exp \left \{- \frac{ z^2}{2(d_z^2/\ln 2)}  \right \},
\label{Z}
\\
&&~~\Phi_{\Lambda_z} (\phi)=\frac{e^{i\Lambda_z \phi}}{\sqrt{2\pi}}.
\label{Phi}
\end{eqnarray}
}
\label{wave}
\end{subequations}
In Eq.\ (\ref{R}), the parameter $R$ describes approximately the
position of the peak of the wave function in the radial
coordinate\footnote{Note that the symbol $R$ here in the microscopic
  description in Eq.\ (\ref{R}) and the same symbol $R$ in the
  phenomenological description in Eq.\ (\ref{eq3}) represent different
  measures of the major toroidal radius and are different physical
  quantities, which can be easily distinguished by the context. }.
The $\rho^{\Lambda}$ dependence of ${\cal R}_\lambda (\rho)$ in
Eq.\ (\ref{R}) arises from the behavior of the wave function near the
origin at $\rho\to 0$, as inferred from the single-particle wave
equation (\ref{eq}).  The quantities $d_{\rho}$ and $d_z$ are related
to the width of the wave function along the $\rho$ and $z$ direction
on the meridian plane.  The normalization constants $N_Z$ and
$N_\Lambda(R,d_\rho)$ are given by
\begin{subequations}
\begin{eqnarray}
&&N_Z=\frac{1} { (2\pi d_z ^2/2\ln 2)^{1/4}},
\\
&&N_\Lambda(R,d_\rho)=\frac{1}{\sqrt{I^{(2\Lambda+1)}(t_{0})}},
\end{eqnarray}
\end{subequations}
where 
\begin{eqnarray}
I^{(2\Lambda+1)}(t_{0}) &\equiv&
 \int_0^\infty     d\rho   ~\rho^{2\Lambda+1}  \exp \left \{- \frac{(\rho-R)^2}{(d_\rho^2/\ln 2)} \right \}~~~~~~~~~~
\\
&=&\!
\left (\frac{d_\rho}{\sqrt{\ln 2}} \right )^{\!\!2\Lambda+2}\!\! \!\frac{\sqrt{\pi}}{2} (2\Lambda+1)! ~i^{(2\Lambda+1)}{\rm erfc}(t_{0}),
\label{ierfc}
\nonumber
\end{eqnarray}
and $i^{(2\Lambda+1)}{\rm erfc}(t_{0})$ is the integral of the error function as defined in Eq. (7.2.3) of \cite{Abr65},
\begin{subequations}
\begin{eqnarray}
i^{(2\Lambda+1)}\text{erfc} (t_{0})&& =\frac{2}{\sqrt{\pi} } \int _{t_{0} }^\infty \frac{(t-t_{0})^{(2\Lambda+1)}}{n!} e^{-t^2} dt  ~~~
\\
\hspace*{-1.1cm}\text{and~~~~~~~~~~~~~~~~~}t_{0}&&=-\frac{R }{d_\rho/\sqrt{\ln 2}}.
\end{eqnarray}
\end{subequations}
The wave functions $\Psi_{\Lambda_z \Omega_z}(\rho,z,\phi)$ in
Eq.\ (\ref{Psi}) for different quantum numbers
$\{\Lambda_z,\Omega_z\}$ are orthonormal.

Instead of the $(d_\rho,d_z)$ parameters, it is convenient to
introduce their geometrical mean, $d$=$\sqrt{d_z d_\rho}$, and the
dimensionless deformation parameter $a_2$ to write them as
\begin{subequations}
{ 
\begin{eqnarray}
d_z&&= d e^{+ a_2 },
\\
d_\rho&&= d e^{ -a_2 }.
\end{eqnarray}
}
\end{subequations} 
The set of $(R,d,a_2)$ parameters give rise to nuclear equidensity
surfaces of different shapes:
\begin{eqnarray}
  R&&=0,    ~a_2=0~:\text{spherical surface},
\nonumber\\
  R&&= 0,     ~a_2>0~ : \text{prolate spheroid},
\nonumber\\
   R&&=0,    ~ a_2<0~: \text{oblate  spheroid},
\nonumber\\
    R&&> 0, ~:\text{spheroid, bi-concave disk,  toroid,}
\nonumber\\
    R&&< 0, ~a_2 < 0 :\text{oblate surface},
\nonumber\\
    R&&< 0, ~a_2 >0 :\text{prolate surface}. 
\nonumber
\end{eqnarray}

\section{$^{12}$C Ground State   Energy  and Density Distribution  }

We construct a Slater determinant for the occupied states and utilize
the Skyrme SkM* interaction \cite{Bar82} to obtain the energy of the
system as a function of the variation parameters $(R,d,a_2)$.  With
the variational wave functions Eqs.\ (\ref{Psi})-(\ref{Phi}) for
neutrons and protons, we get the nuclear density $n(\rho,z)$
\begin{subequations}
\begin{eqnarray}
n(\rho,z)&&=f_\phi f_\rho f_z,
\label{n}
\label{den}
\\
\hspace*{-1.0cm}\text{where}~~~~~~~~~~~~~~~~~~~~~~
f_\phi&&=\frac{1}{2\pi} ,
\\
f_\rho(\rho)&&=4  |  {\cal   R}_0 (\rho)|^2 +8  | {\cal R}_1(\rho)|^2,
\label{frho}
\\
f_z(z)&&= |Z(z)|^2.
\end{eqnarray}
\end{subequations}
The kinetic energy density $\tau (\rho,z)$ defined as \cite{Vau72,Vau73}
\begin{eqnarray}
\tau (\rho,z)=\sum_{\rm occ ~states} |\nabla \psi_{\Lambda_z \Omega_z} (\rho,z)|^2,
\nonumber
\end{eqnarray}
is therefore

\begin{eqnarray}
\tau(\rho,z)=&& f_\phi f_\rho(\rho) |\nabla_z Z(z)\cdot 
\nabla_z Z(z)|
\nonumber\\
&&+f_\phi f_z(z)
\biggl   \{
 4|\nabla_\rho {\cal R}_0(\rho)|^2
+ 8|\nabla_\rho {\cal  R}_1(\rho)|^2
\nonumber\\
&&~~~~~~~~~~+ 8 \frac{\Lambda_z^2}{\rho^2} |{\cal R}_1(\rho)|^2  
\biggr  \}.
\end{eqnarray}
The expectation value of the energy $E(R,d,a_2)$ of the $^{12}$C nucleus in the
mean-field approximation is therefore \cite{Vau72,Vau73}
\begin{eqnarray}
E=\langle H \rangle
&&=\int 2\pi \rho \,d\rho \,dz \,\tau(\rho,z)
\nonumber\\
&&+\int 2\pi \rho \,d\rho \,dz ~\biggl  \{ \frac{3}{8}t_{0}n^2+\frac{1}{16}t_3 n^{\alpha+2} 
\nonumber\\
&&~~~~~~~~+ \frac{1}{16}(3t_1+5t_2)  n \tau
+\frac{1}{64}(9t_1-5t_2)|\nabla n|^2 \biggr  \}
\nonumber\\
&&+\frac{1}{2}\int d^3r_1 d^3 r_2 n({\bb r}_1)  n({\bb r}_2) \frac{e^2}{|{\bb r}_1-{\bb r}_2|},
\end{eqnarray}
where the Skyrme SkM* parameters are
$t_0$=-2645 MeV$\cdot$fm$^3$, $t_1$=410 MeV$\cdot$fm$^5$, 
$t_2$=-135 MeV$\cdot$fm$^5$, $t_3$=15595 MeV$\cdot$fm$^6$, and 
$\alpha$=0.167 \cite{Bar82}.

With the quantities $n(\rho,z)$ and $\tau(\rho,z)$ given in terms of the wave functions, 
which are explicit functions of the variational parameters,
the energy surface $E(R,d,a_2)$ is a function of the variational parameters.
We first study the landscape of the energy surface $E(R,d,a_2)$ of $^{12}$C 
among the nuclear shapes of the sphere, prolate spheroid, and oblate
spheroid.  This can be achieved by constraining $R$ to be zero and
making variations in $d$ and $a_2$.  The contours of the energy
surface $E(R,d,a_2)|_{R=0}$ on the $(d,a_2)$ plane are given in Fig.\ \ref{gsda2}.
\begin{figure} [h]
\hspace*{-0.4cm}
\includegraphics[scale=0.43]{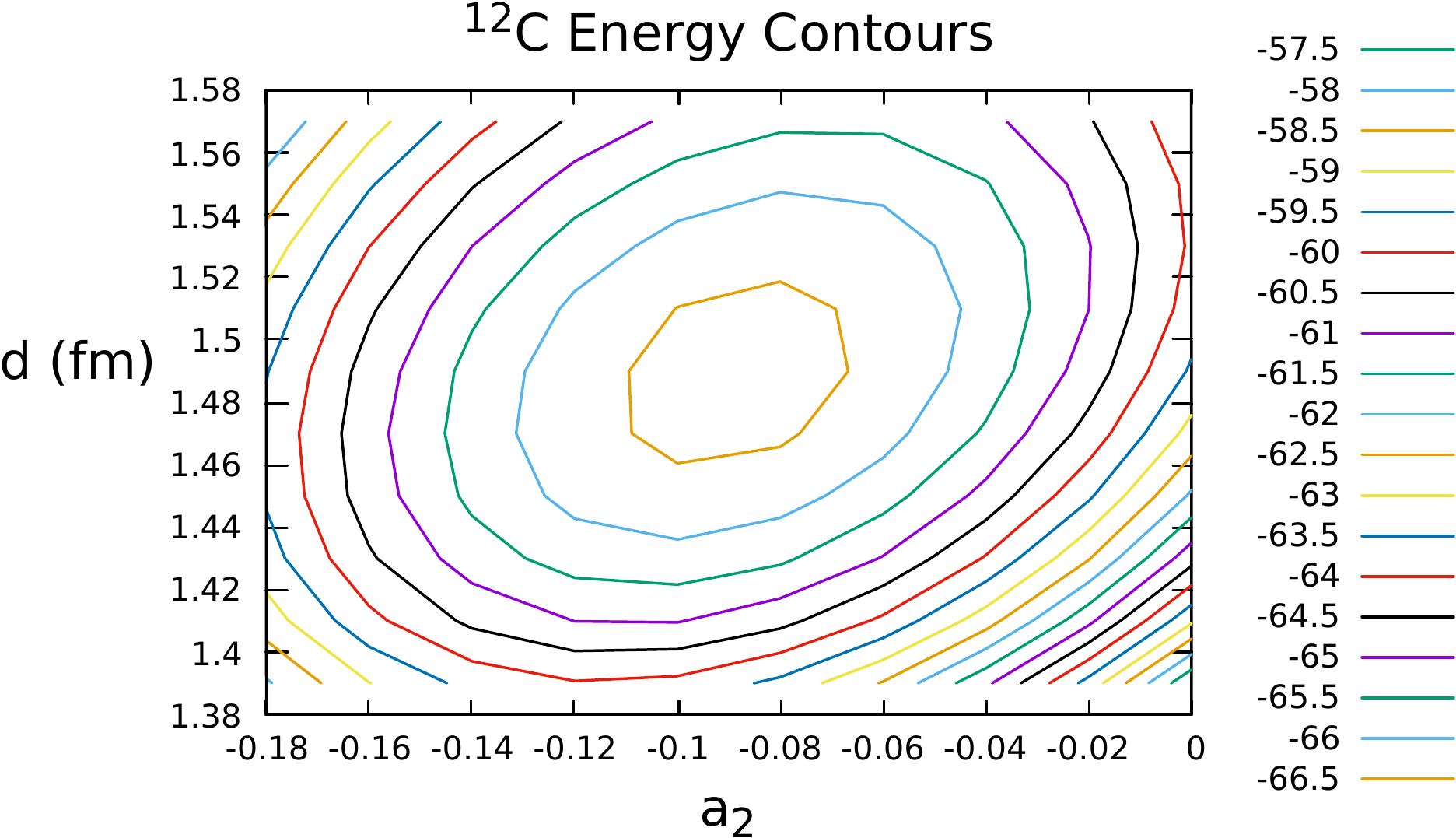}
\caption{(color online). The contours of the $^{12}$C energy surface $E(R,d,a_2)|_{R=0}$
  for $R=$0 on the $(d,a_2)$ plane.  The ground state energy minimum
  appears at $d$=1.49 fm and $a_2$=-0.089 at an energy
  $E$=$E_0$=$-$66.68 MeV.  The energy contours values labeled on the right are in units of MeV.}
\label{gsda2}
\end{figure}

As inferred from Fig.\ \ref{gsda2} for $R=0$, the energy minima at
$d=$ 1.49 fm and $a_2$=$-$0.089 occurs at
$E_{R=0}$$\equiv$$E_0$=$-$66.68 MeV.  The contribution of the
spin-orbit interaction to the ground state of $^{12}$C is $-$23.56 MeV
from self-consistent Hartree-Fock calculations \cite{Sta19}.  If the
spin-orbit interaction is included, the total energy of the system
will be -90.1 MeV which is close to the experimental value of -92.16
MeV.

\begin{figure} [h]
\includegraphics[scale=0.95]{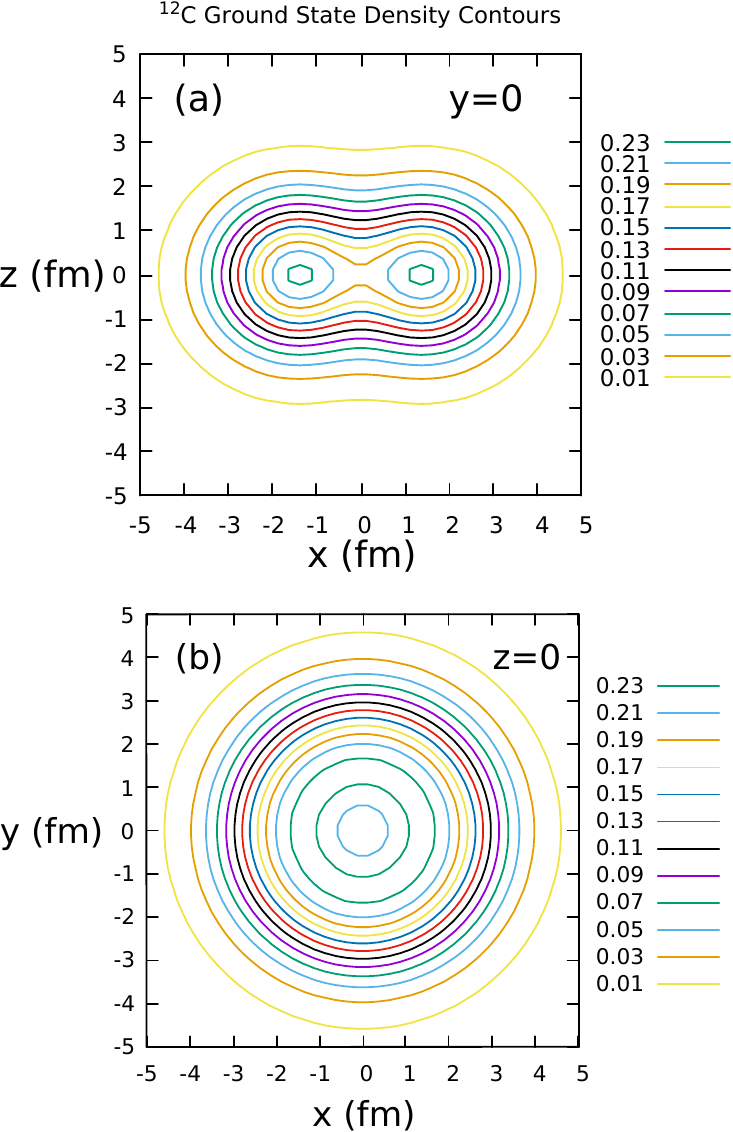}
%\hspace*{-0.cm}
%\includegraphics[scale=0.55]{coul/xzdenR00A}
%
%\includegraphics[scale=0.52]{coul/xydenR00A}
\caption{(color online). The nuclear density  of the $^{12}$C ground state
 (a) on the $y=0$ plane, and (b) on the $z$=0 plane.   The density contours  labeled on the right are  in units of nucleon/fm$^3$.}
\label{fig8}
\end{figure}

We can examine the ground state nuclear density as shown in
Fig.\ (\ref{fig8}a) on the $y$=0 plane, and in Fig.\ (\ref{fig8}b) on
the $z$=0 plane.  We note from Fig.\ (\ref{fig8}a) that the shape of
an equidensity surface of the $^{12}$C ground state depends on the
density value.  The equidensity surfaces in the low density region are
nearly oblate ellipsoids.  The ratio of the axis length along the
$\rho$-direction to the axis length in the $z$-direction at the
half-density surface is approximately 2:1, as would be expected for
the doubly magic shell of $N=6$ and $Z=6$ in an anisotropic harmonic
oscillator potential \cite{Won72}, in agreement with the experimental
deformation of $\beta_2$=$-$0.6 deduced in the direct reaction in
neutron scatterings \cite{Gri67}.  As the density increases, the
equidensity surfaces turn into a bi-concave disks with a small
indentation in the polar regions.  When the density reaches $n \ge
0.21$ /fm$^3$, the equidensity surfaces turn into toroids.  Upon
defining a toroidal nucleus as one in which some equidensity surfaces
are toroidal, the ground state of the $^{12}$C nucleus can be called a
toroidal nucleus.  It has a dense toroidal core immersed in 
lower-density oblate ellipsoids on the surface.

\begin{figure} [h]
\includegraphics[scale=1.10]{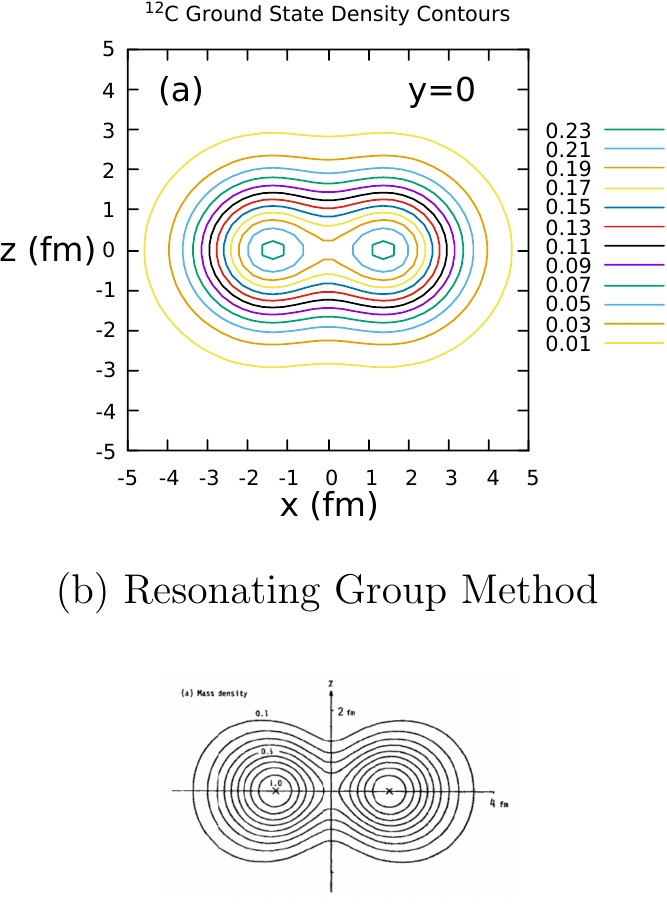}
%\hspace*{-0.7cm}
%\includegraphics[scale=0.50]{coul/xzdenR00A}
%
%\vspace*{0.6cm}
%{  (b) Resonating Group Method }
%
%\hspace*{-1.2cm}\vspace*{0.5cm}
%\includegraphics[scale=0.197]{C12RGMMassDenA}
\caption{(a) The nuclear density of the $^{12}$C ground state on the
  $y$=0 plane obtained by minimizing the energy in the variations of
  $(R,d,a_2)$. The density contours are labeled  in units of nucleon/fm$^3$. 
The density contours  labeled on the right are  in units of nucleon/fm$^3$.
(b) The nuclear density $n$ of the $^{12}$C ground state
  on the $y$=0 plane obtained by the resonating group method (Fig.\ 4
  of Kamimura \cite{Kam81}).
The density contours are labeled  in units of 
1/10 of the
  maximum density.}
\label{fig9}
\end{figure} 

It should be pointed out that the toroidal feature of the $^{12}$C
nucleus of the ground state as exhibited in Figs.\ \ref{fig8}(a) and
\ref{fig9}(a) may appear surprising, but it is in fact in agreement with
earlier results  from the resonating group method presented in Fig.\ 4 of Kamimura
\cite{Kam81} and reproduced in our Fig.\ \ref{fig9}(b), where the density
contours are given in units of  $1/10$ of the  maximum nuclear density. In the
language of the generator coordinates and resonating groups, the
toroidal density distribution in Fig.\ \ref{fig9}(b) arises from a
generator coordinate superposition of the orientation of Wheeler's
triangular cluster wave functions on the cluster plane that generates
naturally an intrinsic toroidal density.  In the language of the
nuclear shell effects \cite{Bra73,Won72,Won73}, the toroidal feature
of the inner core in Fig.\ \ref{fig9}(a) arises from the variation of
the shape of the single-particle wave functions to reach the state of
the lowest energy, and the lowest energy is obtained by settling onto
the doubly-closed toroidal shells of $N $=$Z$=6 \cite{Won72,Won73} resulting in a toroidal density.
From such a perspective,
 there appears to be  a high degree of equivalence and complemntarity 
between the toroidal mean-field description emphasized here and
Wheeler's  triangular 3$\alpha$ cluster description 
investigated elsewhere  \cite{Bri66,AR07,AR07a,AR08,AR09,Hil53,Ueg77,Ueg79,Kam72,Kam74,Hor74,Fun05,Fun09,
  Des12,Kam81}.  

In VAP (Variation after Projection) calculations  in the triangular 3$\alpha$ cluster models
for the ground state and the Hoyle state,
 the variation of the wave function is 
carried out after 
projecting out the $I$=0 and $M$=0 state from the triangular 3$\alpha$ cluster. 
The projection involves the superpositioning of the orientations of the  triangular 3$\alpha$ clusters
on the triangular plane, and the intrinsic nuclear density after projection 
is in essence a toroidal density,
as the results of \cite{Kam81} in Fig.\ \ref{fig9}(b) demonstrate.
The variation after the projection is therefore concerned with the stability of the nucleus 
under the variation of an intrinsic toroidal density.
In  VAP calculations for the $^{12}$C ground state in the resonating group method \cite{Kam81}, the solution of the Wheeler's triangular cluster is
compact, and the corresponding toroidal dense core is immersed in  lower-density oblate
spheroids. In VAP calculations for the Hoyle state,
the AMD and FMD solutions of Wheeler's triangular cluster  have a greater spatial extension
\cite{Oer96,Kan98,Kan01,Kan07,Rot04,Che07,Fel16}, as is also indicated by experimental inelastic scattering experiments 
\cite{Den09}.  It is reasonable to expect a  toroidal nucleus with a more prominent toroidal features and 
a larger toroidal major radius for the Hoyle state.

\section{ Adiabatic  Energy $E(R)$  above the Ground State }

To study the properties of the low-lying excited states above the
ground state, we examine the energy surface and the nuclear density
distributions for different values of $R$ around the ground state of
$^{12}$C.  Under the constraint of a major radius parameter $R$, we
look for the lowest ``adiabatic" energy $E(R)$ by searching for the
$(d,a_2)$ values that lead to an energy minimum.  For a constrained
value of $R$ in question, the adiabatic energy $E(R)$ is determined by
the minimum energy on the $(d,a_2)$ plane.

\begin{figure} [h]
\includegraphics[scale=0.45]{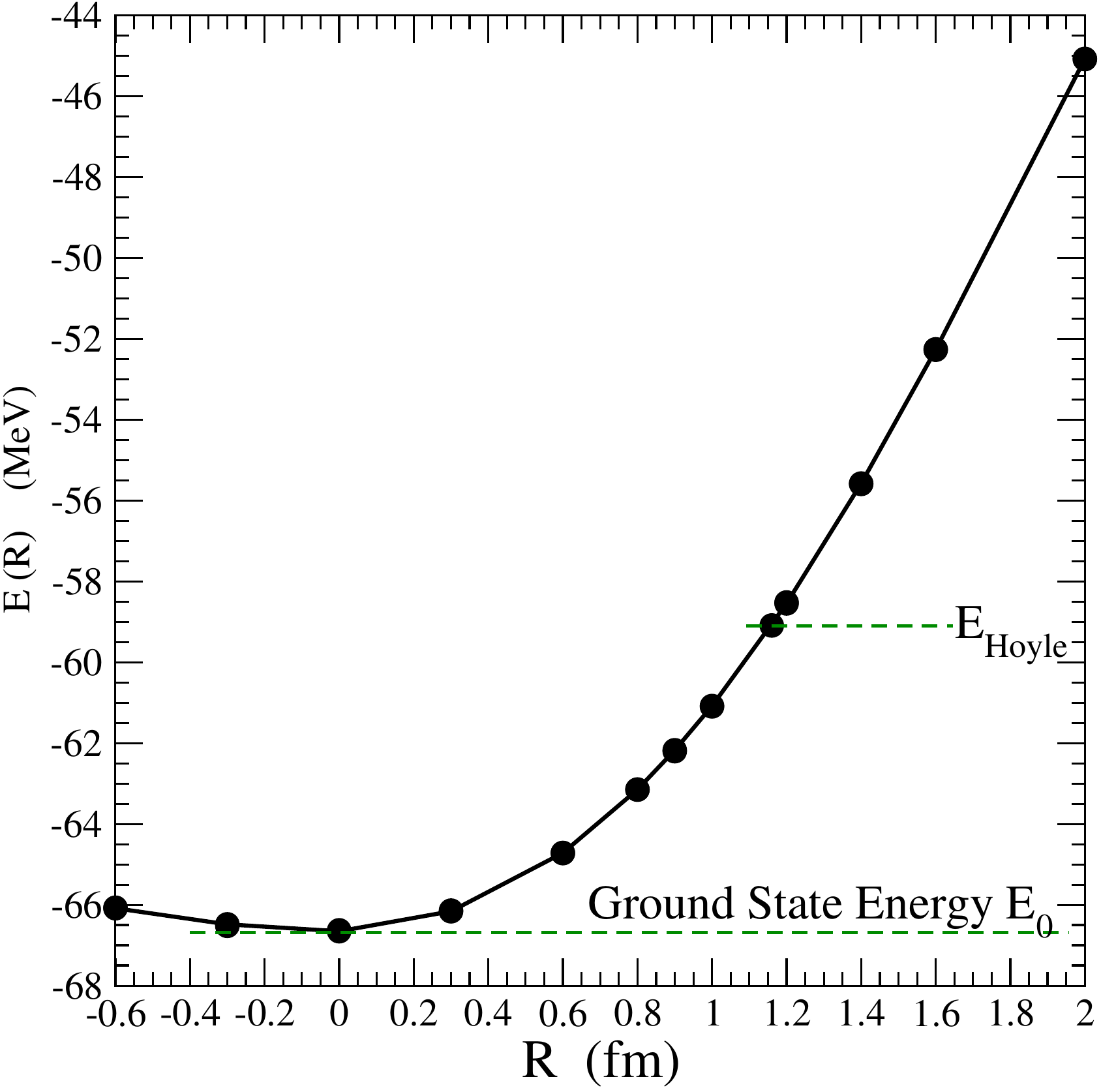}
\caption{ (color online).  The solid points and the adjoining solid
  line give the adiabatic energy $E(R)$ of $^{12}$C.  For each value
  of $R$, the energy $E(R)$ is determined as the minimum of the energy
  in the $(d, a_2)$ plane, whose location at the minimum is listed in
  Table \ref{tbadi}.  }
\label{fig10}
\end{figure}

Figure \ref{fig10} gives the adiabatic energy $E(R)$ as a function of
$R$ and Table \ref{tbadi} lists the corresponding equilibrium location
$(d,a_2)$ at the energy minimum and the excitation energy
$E_x(R)$=$E(R)-E_0$, where $E_0$=$E_{R=0}$. As $R$ increases, the
equilibrium minor radius $d$ decreases, and the equilibrium
deformation parameter $a_2$ increases.  The meridian cross section
changes from an oblate shape to a nearly circular shape on the
meridian plane, as $R$ increases.  In the region of negative value of
$R$, the adiabatic system have nearly the same shape as the oblate
nucleus at $R$=0.  Hence the adiabatic energy flattens out in the
region of negative $R$.

We find in Table \ref{tbadi} and Fig.\ \ref{fig10} that the ground
state energy $E_0$ at $R=0$ is a minimum not only under the variation
of $(d,a_2)$ but also a minimum under a variation of $R$.
Fig.\ \ref{fig10} shows further that as a function of $R$ there is no
additional local energy minimum at a non-zero positive value of $R$ in
the mean-field approximation.

\begin{table}[h]
\caption { For each value of $R$, the energy $E(R)$ is the minimum of
  the energy on the $(d, a_2)$ plane, for which the $(d,a_2)$ location
  at the minimum are listed.  The excitation energy $E_x$ is the
  energy $E(R)$ relative to the ground state $E_0$,
  $E_x$=$E(R)$-$E_0$, where $E_0$=$E_{{}_{R=0}}$.  }
\vspace*{0.3cm}

\centering
\begin{tabular}{|c|c|c|c|c|}
\hline
$R$   &  $E(R)$  &    $E_x$  & $d$  & $a_2$    \\
(fm)  &  (MeV) &     (MeV) &  (fm) &   \\
  \hline
 -0.60  & -66.07     &    0.61 & 1.570 &  -0.140  \\
 -0.30  &   -66.52   &    0.16 & 1.540 &  -0.120\\
  0.0    &   -66.68   &    0.00   & 1.493 &  -0.089\\
  0.3    &  -66.14    &    0.54   & 1.450 &  -0.065 \\
  0.6    &  -64.72    &    1.96  &1.420 &  -0.050\\
  0.9   &    -62.20   &    4.48  & 1.394 &  -0.025\\
  1.0   &   -61.09    &    5.59 & 1.377 &  -0.020\\
 1.16  &  -59.03   &     7.65  &  1.370&  -0.007   \\
  1.2   &   -58.52    &    8.16 & 1.368 &  -0.004\\
  1.4   &   -55.59    &    11.09 & 1.360 &   0.004\\
  1.6   &   -52.26    &    14.42 & 1.354 &   0.010\\
   2.0  &  -45.08     &     21.60 & 1.350  &   0.020 \\
 \hline
\end{tabular}
\label{tbadi}
\end{table}
\vspace*{0.3cm}

We can examine the energy surface in other degrees of freedom in order
to search for a secondary local energy minimum.  The toroidal shape
density can evolve into a 3$\alpha$ cluster distribution by the
sausage deformation $\sigma_3$ of order $\lambda=3$, which turns an
azimuthal-symmetric toroid into three clusters with three adjoining
necks, (as illustrated in Fig. 3 of Ref.\ \cite{Won73}).  To study the
question of stability with respect to a variation in $\sigma_3$, we
modify the wave function $\Phi_{\Lambda_z}(\phi)$ of Eq.\ (\ref{Phi})
to
\begin{eqnarray}
\Phi_{\Lambda_z}(\phi) = \frac{ (1+\sigma_3 \rho^3 e^{3i \phi}) \rho^\Lambda e^{i \Lambda_z \phi} }{\sqrt{2\pi}}.
\end{eqnarray}
Direct calculations in the mean-field theory with Skyrme SkM*
interactions reveal that in the neighborhood of the Hoyle energy
region the toroidal state, $\Psi_{\rm toroid}$, the nucleus system is
stable against sausage deformation of order $\lambda$=3 at
$\sigma_3$=$\sigma_{3T}$=0.  There is no local energy minimum at a
non-zero value of $\sigma_3$ and $R$ as shown in Fig.\ \ref{Rb3d1368}.
 \begin{figure} [h]
\hspace*{-0.1cm}
\includegraphics[scale=0.47]{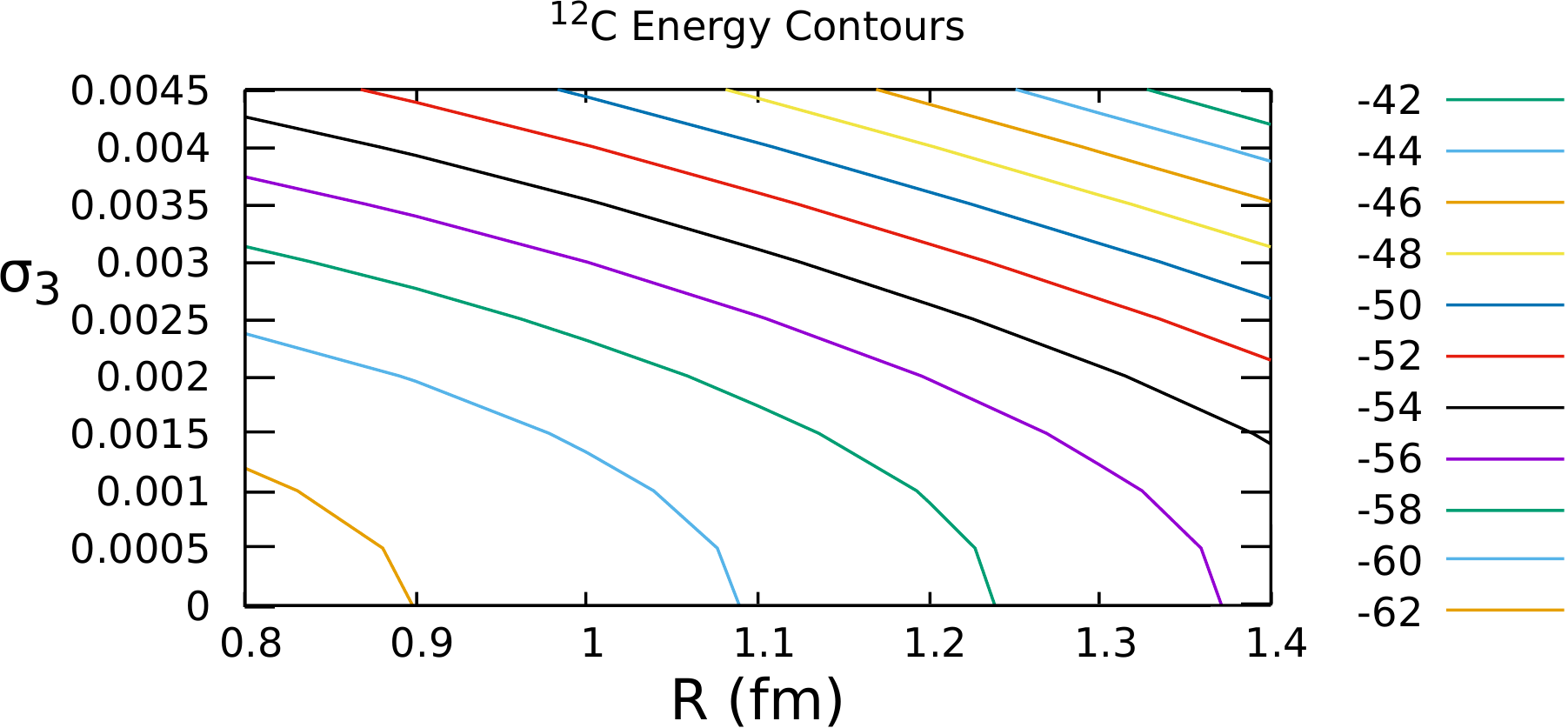} 
\caption{(color online). The energy contours  
of the energy surface $E(R,\sigma_3)$ in the plane of $(R,\sigma_3)$ for the set of parameters $d$=1.370 fm  and $a_2$=-0.007 appropriate for the excitation energy at the Hoyle energy.  The energy contours are labeled in units of MeV.}
\label{Rb3d1368}
\end{figure}
We can further explore different $R_\Lambda$ parameters for the
different $\Lambda$ single-particle states.  Upon considering the
parameters space of $(R_{\Lambda=0},R_{\Lambda=1},d,a_2,\sigma_3)$, we
find that in the mean field approximation with the Skyrme SkM*
interaction, there is only a single ground state energy minimum at
$(R_{\Lambda=0},R_{\Lambda=1},d,a_2,\sigma_3)$=(0, 0, 1.493fm, -0.089,
0), with no secondary local energy minimum in the neighborhood of the
Hoyle energy.  Extensive Hartree-Fock and Hartree-Fock-Bogoliubov
calculations using the Skyrme SkM* interactions also fails to indicate
a secondary toroidal local energy minimum in the neighborhood of the
Hoyle energy \cite{Sta19}.

\section{Density Distribution of the $^{12}$C Nucleus at Hoyle Energy}

\subsection{Toroidal State $\Psi_{\rm toroid}$ at Hoyle  Energy}

\begin{figure} [h]
\includegraphics[scale=0.90]{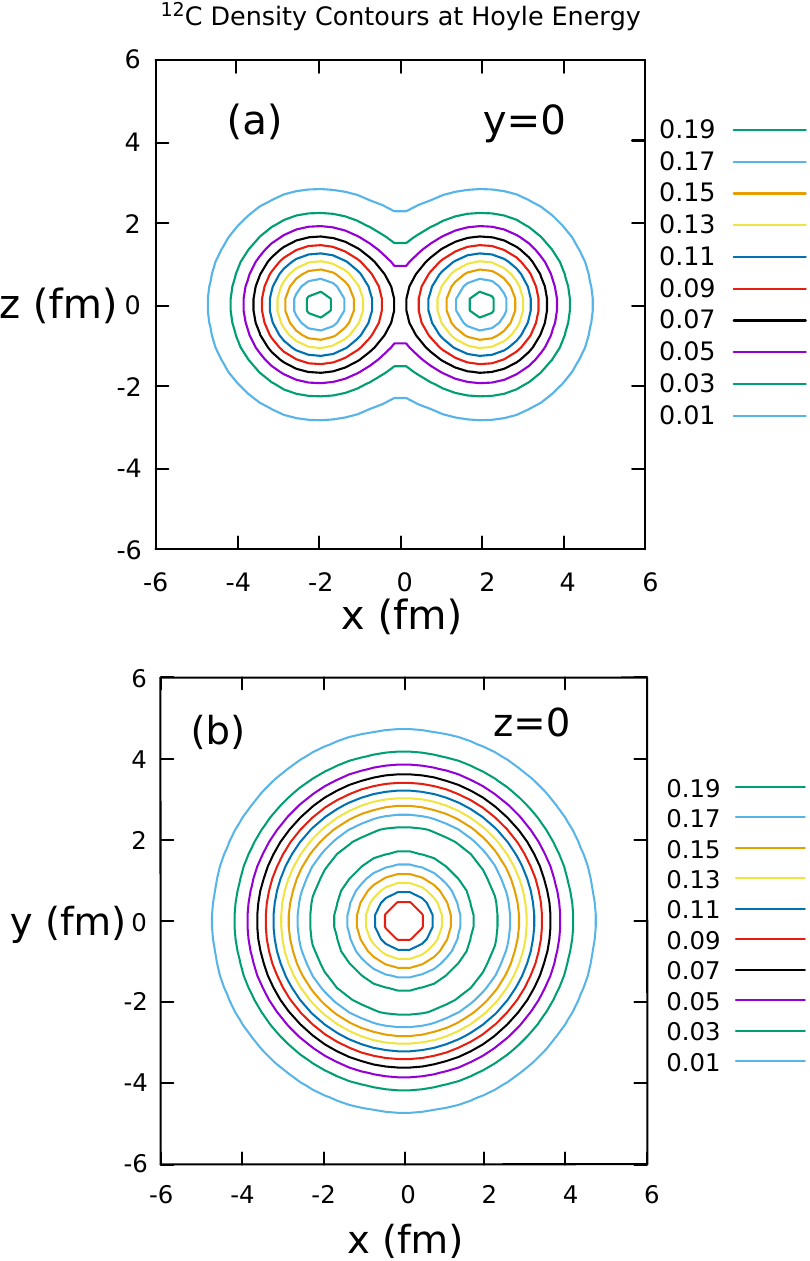}
%\hspace*{-0.1cm}
%\includegraphics[scale=0.61]{coul/xzdenR116A}
%\hspace*{-0.1cm}
%\includegraphics[scale=0.585]{coul/xydenR116A} 
\caption{(color online). The nuclear density for the toroidal state of
  $^{12}$C at the Hoyle state excitation energy, (a) on the $z$=0
  plane, and (b) on the $y$ plane.  The density contours are labeled  in units of nucleon/fm$^3$.}
\label{xydenR116A}
\end{figure}

The adiabatic energy $E(R)$ examined in Fig.\ \ref{fig10} and listed
in Table \ref{tbadi} allows us to find the variational parameter
values ($R$, $d$, $a_2)$ at which the $^{12}$C excitation energy is at
the Hoyle excitation energy.  From Table \ref{tbadi}, we find that the
case of ($R$=1.16 fm, $d$=1.370 fm, $a_2$=-0.007) corresponds to an
excitation energy of the Hoyle energy at $E_x$=7.65 MeV.  With this
set of variational parameters, we can calculate the nuclear density
distribution of the state of the system at the Hoyle energy.  We plot
the nuclear density on the $x$-$z$ plane at $y$=0 in
Fig.\ \ref{xydenR116A}(a) and the density contours on the $x$-$y$
plane at $z$=0 in Fig.\ \ref{xydenR116A}(b).  One observes that for
this state of the system at the Hoyle energy on the $E(R)$ energy
surface, the equidensity surfaces with density $n\ge 0.07$/fm$^3$
appear as separated toroids, while those with lower densities as
spindle toroids.  The meridian cross sections are nearly circular.
Thus, the toroidal core that shows up for the $^{12}$C ground state at
$R$=0 becomes much more prominently toroidal, as $R$ increases to 
$R$=1.16 fm at the
Hoyle energy region.  One concludes from such an investigation that
there is a state at the Hoyle energy with a prominent toroidal density
distribution shown in Fig.\ \ref{xydenR116A} that is degenerate with
the energy of the Hoyle state.  It is a Slater determinant consisting
of single-particles states of Eqs.\ (\ref{Psi}-\ref{Phi}) with the
appropriate $R$, $d$, and $a_2$ values.  In the mean-field theory,
such a state is stable against variations in $d$ and $a_2$ but lies on
an energy slope as a function of $E(R)$, as shown in Fig.\ \ref{fig10}.  It is therefore unstable
against the contraction of the major radius $R$ in the mean-field
approximation.  We nonetheless call it a provisional toroidal state
at the Hoyle energy, $\Psi_{\rm toroid}$, on account of its toroidal
density shape, pending further investigation of its stability against
$R$ variations.  We shall examine whether many-body interactions
beyond the mean field may affect the stability of the provisional  toroidal state
$\Psi_{\rm toroid}$ against variations in $R$.  It should however be
kept in mind that pending modifications arising from additional
interactions beyond the mean field may modify slightly the $R$
location and the shape of the density distribution but will not likely
change its toroidal characteristics.

\begin{figure} [h]
\includegraphics[scale=0.35]{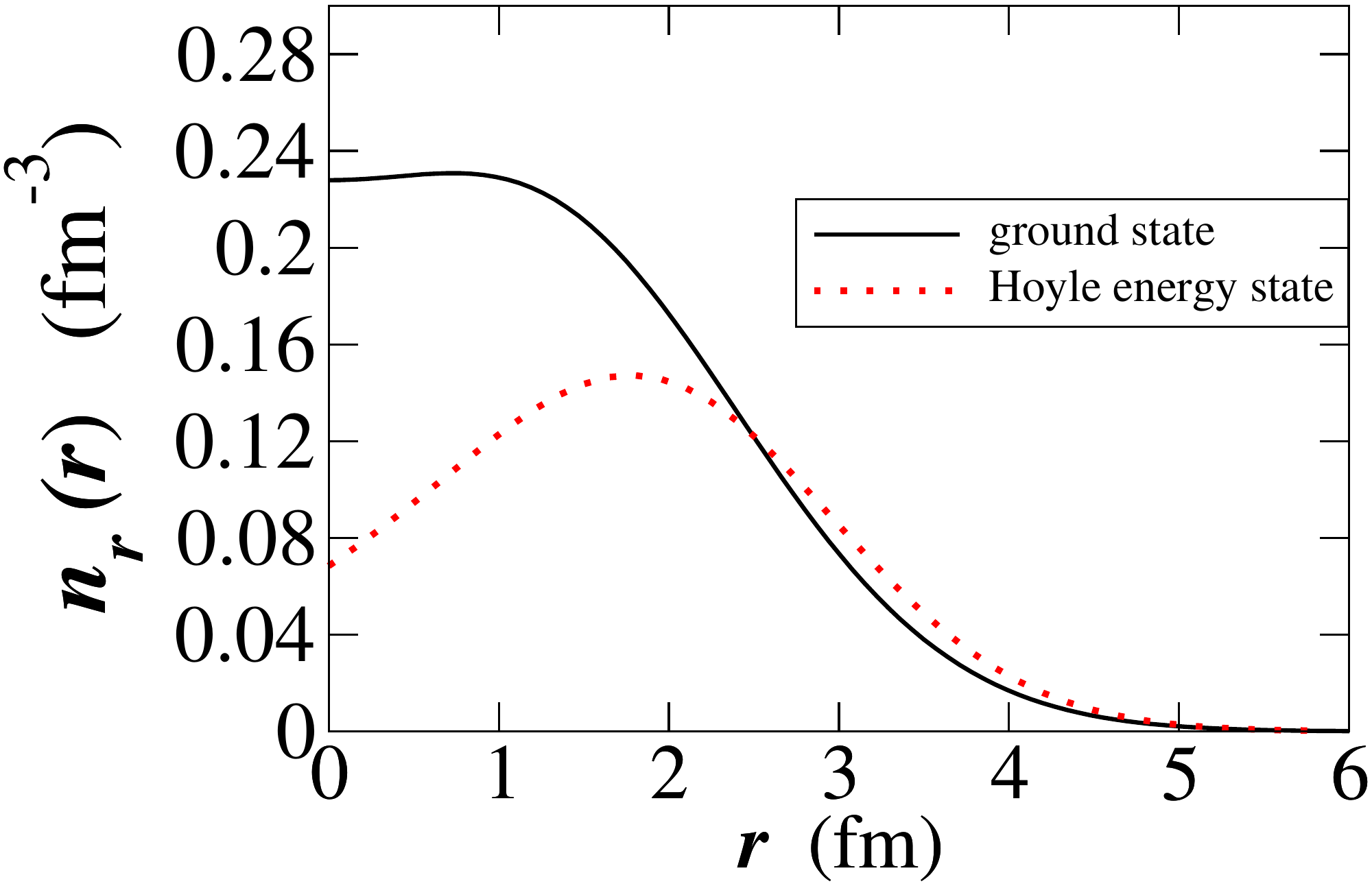}
\caption{The nuclear density $\rho(r)$ in the spherical coordinate
  system, after integrating over the polar angles. (a) Results from
  the antisymmetric molecular dynamics calculations of \cite{Kan07},
  and (b) from the provisional toroidal state at $R=1.16$ fm,
  $\Psi_{\rm toroid}$, at the Hoyle energy.  }

\label{rhor}
\end{figure}

We can get another idea on the spatial extension of the ground
 state and the provisional toroidal state $\Psi_{\rm toroid}$
plot the density distribution $n_r(r)$ of $\Psi_{\rm
   toroid}$ as a function of the spherical radial coordinate $r$ after
 integrating over the polar angle $\theta$,
\begin{eqnarray}
n_r(r)= \int_0^\pi n(\rho,z) \sin (\theta) d\theta, 
\end{eqnarray}
for which $2\pi \int n_r(r) r^2 dr= A$.  The angle-integrated densities $n_r(r)$ as a function of the  spherical coordinate $r$ 
 for the ground state (with
 $R=0$) and the provisional toroidal state at Hoyle energy (with $R$=1.16
 fm) are plotted in Fig.\ \ref{rhor} .   We note that  the density $n_r(r)$  in the interior
 of the ground state is about two to three times that of the provisional toroidal state $\Psi_{\rm toroid}$,
but the latter has a more extended distribution at large $r$ values.

\subsection{Models of 3$\alpha$ Clusters}

Even though we do not find a toroidal local energy minimum as a
function of $R$ in the mean-field theories, we are however motivated
to continue the search in view of the many pieces of experimental
evidence supporting the tentative identification the Hoyle state and
many of its excited states to be states of $^{12}$C in a toroidal
configuration, as discussed in Section V. From intuitive viewpoints,
we are further encouraged by the small energy separation between the
Hoyle state and its excited states, by the large number of both the
identified and the un-identified broad excited states, by the close average
energy spacing between the states, and by their predominance in their
decay into three alpha particles. These characteristics suggest that
the Hoyle state is intrinsically a spatially extended object that is
capable of possessing a complex particle-hole excitation structure. A
toroidal description is consistent with such a
suggestion. Theoretically we find it promising that the ground state
nuclear density as exhibited in Fig.\ \ref{fig8} already shows a
toroidal structure in its core and the provisional state at the Hoyle
energy in Fig.\ \ref{xydenR116A} exhibits prominent toroidal
characteristics. We also note with interest that there
may be residual interactions beyond the mean-field that may be
peculiar to $^{12}$C and have important effects on the stability
of the toroidal system as a function of $R$.

We seek guidance from earlier models that have been successful in
describing both the ground state and the Hoyle state. Cluster models
of different types, originating from Wheeler's concept of a triangular
cluster of three alpha particles, have been quite successful in
explaining many salient features of the states of the $^{12}$C
nucleus.  There is the interacting cluster model of three alpha
particles in which the alpha particles are approximated to be
structure-less for simplicity but the dynamics of the clusters are
solved as a quantum mechanical three-body problem
\cite{AR07,AR07a,AR08,AR09}. There are the resonating group method
\cite{Whe37,Whe37a,Kam81} and the generator coordinate method of alpha
clusters \cite{Hil53,Ueg77,Ueg79,Des12} in which the nucleons and
their exchanges are described microscopically and the dynamics between
clusters is determined by a variational principle with the variational
wave function as a superposition of triangular 3$\alpha$ clusters
wave function. The ground state and the Hoyle state appear as
quantized solutions of the Hill-Wheeler equation appropriate for a
system of three alpha particles with different radial extensions. In
addition, there are also the AMD model \cite{Oer96,Kan98,Kan01,Kan07}
and the FMD model \cite{Rot04,Che07,Fel16} in which the cluster states and
the shell model states coexist, and the nucleons cluster automatically
by themselves upon a minimization of the energy of the system.

All the above mentioned models find the ground state and the Hoyle
state as 3$\alpha$ clusters, with the Hoyle state to be spatially more
extended than the ground state. Although the ground state may be
adequately described by the independent particle mean-field theory,
the description of the Hoyle state in all these descriptions appears
to require effectively interactions beyond the mean field.

We shall discuss the simplest cluster model of
Ref. \cite{AR07,AR07a,AR08,AR09} as a representative, in order to
bring out the most important ingredients. In this cluster model of
$^{12}$C as a quantum-mechanical three-body problem with structureless
alpha particles, the two-alpha interaction is taken from a
phenomenological Ali-Bodmer interaction that describes well the
properties of the two-alpha systems \cite{Ali66}.  It was found that
good agreement of the properties of the 
low-lying  states
and their decay widths \cite{AR07,AR07a,AR08,AR09} necessitates the introduction of an attractive
three-alpha cluster  interaction taken to be of the form

\begin{eqnarray}
&&V_{3 \alpha}({\bb r}_1 {\bb r}_2 {\bb r}_3) = S \exp\{ -\rho^2/b^2\},
\label{3bb}
\end{eqnarray}
where ${ \bb r}_i$ is the coordinate of the $i$th alpha particle and
\begin{eqnarray}
\rho^2=\frac{4}{3} \sum_{i<j}^3 ({\bb r}_i - {\bb r}_j)^2/b^2\}.
\end{eqnarray}
The range parameter $b$ has been taken to be 6 fm in
\cite{AR07,AR07a,AR08,AR09}, corresponding to the situation that the
case of $\rho$=$b$ corresponds to the condition of three touching
alpha particles. For the Hoyle state, $S $ was found to be $-$20
MeV. The occurrence of the Hoyle state in the collision of three alpha
particles in stellar evolution, leading to the nucleosynthesis from
light elements to heavy elements, provides an additional strong
support for the presence of this type of three-alpha cluster residual
interaction.

In a completely microscopic picture including the constituents of the
alpha particles, the three-alpha interaction such as Eq.\ (\ref{3bb})
will involve all twelve nucleons that is beyond the scope of the
mean-field approximation.

\subsection{ Possible Coexistence of the Toroidal  and the Three-alpha Cluster
Configurations}

With the addition of the 3$\alpha$ cluster force, the quantum-
mechanical three-body problem was solved to obtain the 3$\alpha$
cluster state $\Psi_{3\alpha}$ that gives a good description of the
Hoyle state energy and decay width as described in
\cite{AR07,AR07a,AR08,AR09}. We have thus the situation that there are
two degenerate states with distinctly different density distributions
but with the same energy : (i) the (provisional) toroidal state
$\Psi_{\rm toroid}$ with a toroidal density distribution at the Hoyle
energy as shown in Fig.\ \ref{xydenR116A}, and (ii) the $3\alpha$
cluster state $\Psi_{3\alpha}( {\bb r}_1{\bb r}_2 {\bb r}_3)$ at the
Hoyle energy as obtained in \cite{AR07} by solving the three-body
problem. They are degenerate at the Hoyle energy. Because of the
energy degeneracy, the two states must mix with each other.

\begin{figure} [h]
\hspace*{-0.1cm}
\includegraphics[scale=0.50]{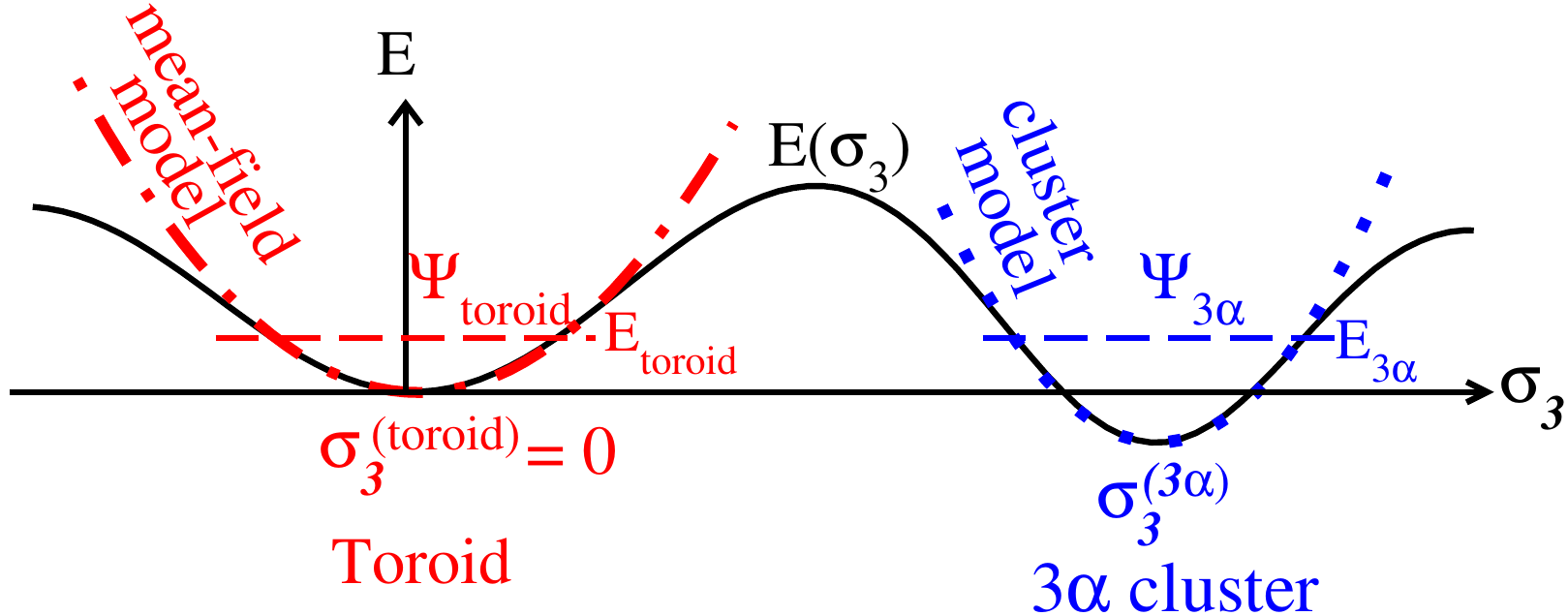}
\caption{(color online).  Schematic depiction of the energy curve  $E(\sigma_3)$ (solid  curve) in
  the sausage degree of freedom, $\sigma_3$ at the Hoyle energy,
  $E_{\rm toroid}$=$E_{3\alpha}$=$E_{\rm Hoyle}$.  The toroidal state $\Psi_{\rm toroid}$
  is stable against $\sigma_3$ variations at $\sigma_{3}$=$\sigma_{3}^{({\rm toroid})}$=0.  The
  3$\alpha$ cluster state $\Psi_{ 3\alpha}$ is stable against
  $\sigma_3$ variations at $\sigma_3$=$\sigma_{3}^{(3\alpha)}$.  The two
  degenerate configurations mix with each other to give rise to a pair
  of states.  }
\label{twowell}
\end{figure}

The toroidal shape density can evolve into a three-$\alpha$ cluster
density by the sausage deformation $\sigma_3$, of order $\lambda=3$.
We can envisage the behavior of the energy surface  $E(\sigma_3)$ 
similar to those in a double-well potential or   a fission isomer \cite{For59,Bra73,Kol75},
as shown schematically in Fig. \ref{twowell}.
Direct calculations in a mean-field theory reveal in Fig.\ \ref{Rb3d1368} that the toroidal
state $\Psi_{\rm toroid}$ at the Hoyle energy is stable against
sausage deformation of order $\lambda=3$ at $\sigma_3=\sigma_{3}^{({\rm toroid})}=0$,
as depicted schematically by the dashed-dot curve in
Fig.\ \ref{twowell}.  On the other hand, by including the essential
three-alpha attractive interaction [Eq.\ (\ref{3bb})] in the quantum
mechanical three-body problem, the 3$\alpha$ cluster model yields a
resonance $\Psi_{3\alpha}$ at the Hoyle energy, with the correct
three-alpha decay width \cite{AR07}. The presence of such a 3$\alpha$
cluster state at the Hoyle energy means that the 3$\alpha$ cluster
state is a local energy minimum at a non-zero value of
$\sigma_3=\sigma_{3}^{(3\alpha)}$, as depicted as the dotted curve in
Fig.\ \ref{twowell}.  Thus, there are two degenerate states,
$\Psi_{\rm toroid}$ and $\Psi_{3\alpha}$, at the Hoyle energy.  A
complete solution will involve the full $\sigma_3$ space with both
shapes to lead to the mixing of the two configurations, similar to the
case of fission isomers \cite{Bra73} and shape isomerism in Hg
isotopes \cite{Kol75}.  A solution for two physical states in the full
$\sigma_3$ space will be of the type
\begin{subequations}
\begin{eqnarray}
\Psi_{\rm I}&&= a_{\rm toroid}  \Psi_{\rm toroid} + a_{3\alpha} \Psi_{3\alpha},
\label{PsiI}
\end{eqnarray}
\end{subequations}
as in a double-well potential and  fission isomers \cite{For59,Bra73,Kol75}.  The two states $
\Psi_{\rm toroid}$ and $\Psi_{3\alpha}$ are not orthogonal as their
overlap $ \langle \Psi_{\rm toroid} | \Psi_{3\alpha}\rangle $ is in
general not zero.  The Hamiltonian matrix for these two states needs
to be constructed and diagonalized to obtain the pair of physical
states.  The splitting between the pair of two physical states
involves a tunneling through the barrier between the the toroidal
state at $\sigma_3$=$\sigma_3^{({\rm toroid})}$=0 to the three-alpha cluster and the local energy
minimum at  a finite $\sigma_{3}^{(3\alpha)}$ \cite{For59}.

\subsection{Effective 3$\alpha$ Cluster Interaction beyond the Mean Field}

The presence of the attractive three-cluster interaction $V_{3\alpha}$
such as that represented in Eq.\ (\ref{3bb}) will lead to a correction
to mean-field adiabatic surface $\Delta E_{3\alpha}(R)$ for the
toroidal configuration. We envisage that the many-nucleon system leads
to a physical state at the Hoyle energy that has a toroidal component
$\Psi_{\rm toroid} (R)$ described by a Slater determinant of the
toroidal single-particle orbitals given in
Eqs.\ (\ref{Psi}-\ref{Phi}).  The effect of a three-alpha particle
cluster interaction $V_{3\alpha}$ leads to a change of the adiabatic
energy in the toroidal sector given by
\begin{eqnarray}
&&\Delta E_{3\alpha}(R) = \langle \Psi_{\rm toroid}(R)|V_{3\alpha}| \Psi_{\rm toroid}(R) \rangle
\nonumber\\
&&~=\!\! \!\int \!\! d {\bb r}_1\!d {\bb r}_2 d {\bb r}_3
|\!\langle \Psi_{\rm toroid}(R)  | \Psi_{3\alpha}({\bb r}_1{\bb r}_2 {\bb r}_3 )\rangle\!|^2
V_{\rm 3\alpha}({\bb r}_1{\bb r}_2 {\bb r}_3).
\nonumber
\label{overlap}
\end{eqnarray}
Here, the three-alpha cluster state $\Psi_{3\alpha}({\bb r}_1{\bb r}_2
{\bb r}_3 )$ is quantized in accordance with the three-body wave
equation \cite{AR07} and it has a definite root-mean-square
radius. Therefore, the probability $|\langle \Psi_{\rm toroid}(R) |
\Psi_{3\alpha}({\bb r}_1{\bb r}_2 {\bb r}_3 )\rangle|^2$ can be
represented by a Gaussian with a width parameter $a_{{}_ R}$ centered
at the $R_0$ value, for which $\Psi_{\rm toroid}( R_0)$ and the
solution $ \Psi_{3\alpha}({\bb r}_1{\bb r}_2 {\bb r}_3 )$ have the
same root-mean-square radius. We are led to a semi-empirical adiabatic
energy correction arising from the 3-alpha clustering interaction of
the form
\begin{eqnarray}
\Delta E_{3\alpha}(R) =  A e^{-(R-R_0)^2/2 a_{{}_R}^2}
\end{eqnarray}
The set of parameters of $A$=$−$6 MeV, $R_0$=1.6 fm, and $a_R$=0.15 fm
will give a secondary toroidal energy minimum at the Hoyle energy of
$E_x$ =7.65 MeV at $R$=1.55 fm. They serve here only as an example to
indicate that there are known 
interactions beyond the mean field that may have important effects on the stability of the 
$^{12}$C nucleus in the toroidal configuration at the Hoyle energy.
 Much more work will need to be 
carried out   to clarify the situation.  The phenomenological to investigate toroidal shape 
should continue to proceed 
because both the microscopic foundation and the phenomenology on the intrinsic shape of the Hoyle state 
will benefit from progress in either direction.

\section{Conclusions and discussions}

In spite of many investigations on the excited states of $^{12}$C, the
physical nature of the Hoyle state and its many excited states remains
an interesting puzzle
\cite{Bec10,Whe14,Whe15,Zim13,Smi17,Del17,Gai17,Kel17,Kir10,Kir12,Alc12,Bar18,Hor10,Fre17,Des12,
  AR07,AR07a,AR08,AR09}.  We explore the toroidal degree of freedom
for the reason that the $^{12}$C nucleus, with 6 neutrons 6 protons,
is a doubly closed-shell nucleus in a toroidal potential.  A generator
coordinate superposition of the orientations of Wheeler's triangular
cluster states on the cluster plane will naturally generate a toroidal
density.  Many excited states of $^{12}$C decays predominantly into
three alpha particles, and a cluster of three alpha particles has a
probability amplitude overlap with the toroidal wave
function. Recently, experimental evidence for toroidal high-spin
isomers in $^{28}$Si predicted by a number of theoretical
investigations have recently been reported \cite{Cao19}. For these
reasons, we study the states of the $^{12}$C nucleus in a toroidal
configuration from both phenomenological and microscopical viewpoints.

In the phenomenological approach, we search for the signature of a
toroidal nucleus and we find that the $^{12}$C single-particle state
energies have a simple $\Lambda^2/R^2$ structure and these states
bunch together to form single-particle $\Lambda$-shells. Consequently
the toroidal shell structure gives rise to particle-hole multiplets of
$^{12}$C excited states bearing the signature for the intrinsic
toroidal properties. The spectrum of a toroidal nucleus is
characterized by these shell-to-shell step-wise particle-hole
excitation multiplets of various spin, parities, and excitation
energies.

Upon comparing with the experimental spectrum, we find approximate
agreement of the spins and parities of identified low-lying
states. The underlying broad structures in the excitation functions of
the $^{11}$B($^{3}$He,d)$^{ 12}$C$^*$$\to$3$\alpha$ reaction of
Ref.\ \cite{Kir10} indicates the possible presence of the remainder
members of the produced multiplets. The
$^{10}$B($^3$He,p)$^{12}$C$\to$3$\alpha$ data at higher energies
indicates possible copious production of toroidal states as a large
underlying broad structure underneath the resolved resonances. 
Subject to further experimental and theoretical investigations, the
Hoyle state and many of its excited states may be tentatively
attributed to be states of a $^{12}$C nucleus in the toroidal
configuration, with the Hoyle state as the band head. There may be a
large number of toroidal $^{12}$C states over a large energy region
that readily breakup into three alpha particles, which may have
implications in energy-producing mechanisms.

In the microscopic mean-field approximation with the Skyrme energy
density functional for the $^{12}$C nucleus, we find that the ground
state of $^{12}$C has an low oblate spheroidal density distribution 
on the surface but a high density  toroidal distribution in the core, in agreement with previous calculations using the generating coordinate method.  At the Hoyle energy, the nuclear density exhibits a pronounced
toroidal structure. On the other hand, previous quantum three-body
treatment of the $^{12}$C nucleus yields the proper resonance energy
and width at the Hoyle energy when a three-body force is introduced
\cite{AR07,AR07a}. Because the toroidal state can evolve into a
three-alpha cluster through the sausage deformation $\sigma_3$ and the
toroidal state and the 3$\alpha$ cluster states have the same energy,
the physical state is likely a mixture of the toroidal state and the
3$\alpha$ cluster state, suggesting the possibility of a toroidal
coexistence of the physical state possessing probability amplitudes
for both the toroidal and the three-alpha cluster configurations. The
physical Hoyle state and many of its excited states may therefore
exhibit toroidal nucleus characteristics with the presence of
particle-hole multiplets examined here as well as three-alpha cluster
characteristics studied in cluster models.  Such a coexistence model
is similar to the coexistence model \cite{Hor74,Hor10} proposed
earlier. The difference is that the cluster model with shell model
wave function is hereby replaced with a toroidal structure that is
geometrical in its content.

It is instructive to compare and contrast the advantages and
disadvantages in the toroidal and the three-alpha cluster descriptions of
the $^{12}$C nucleus under different physical probes in different
measurements. In the toroidal configuration, nucleons traverse
azimuthal orbitals and a particle-hole excitation can be promoted
easily from one azimuthal $\Lambda$-shell to another $\Lambda$-shell
with low expenses in energy, of order a few MeV. On the other hand, in
the three-alpha cluster of strongly bound alpha particles as a dilute
gas, a particle-hole excitation of the alpha particles will require a
very high energy, of order 20 MeV. Hence, the toroidal configuration
may be more
efficient for problems involving particle-hole excitations, as for
example, in the stripping-type experiments carried out in the
Aarhus-Madrid Collaboration in \cite{Kir10,Kir12,Alc12}. In matters of
the isolation of the four correlated particles as an escaping entity
and the tunneling through the eternal Coulomb barrier the three-alpha
cluster configuration may provide a more convenient description. A
toroidal coexistence picture of the $^{12}$C nucleus as a mixing of
both the toroidal and three-alpha cluster configurations may therefore
be a useful concept.

The proposed toroidal description of the $^{12}$C states may be useful
not only to correlate existing experimental data, it has a number of
predictions that will stimulate future experimental and theoretical
work.  Specifically, the toroidal particle-hole picture predicts many
additional states in the particle-hole multiplet, and it is likely
that these are the broad states in the underlying structure under the
identified sharp states, as is discussed in Section V.  Therefore,
experimental identification of the spin and parity of the broad states
in the underlying structure, if at all possible, will be useful.  In this
respect, recent experimental effort in finding the spin and parity of
a broad state in the continuum can be of value \cite {Kir13}.
Theoretically, it will also be of interest to plot the intrinsic
nuclear density on the $x$-$z$ plane for the Hoyle states in the
GCM and RGM calculations or other microscopic calculations.  The
present work would predict that the Hoyle state will show an intrinsic
toroidal density, just as the ground state shows an intrinsic toroidal
dense core in the GCM work with a 3$\alpha$ cluster in the work of
Kamimura \cite{Kam81} discussed in Section 10A.  Theoretically, much
work needs to be done to understand quantitatively the excitation
function to see if the spectroscopic factors for the different states
in the multiplets are indeed as they should be.  We also need to
understand theoretically why some states have narrow width while some
other have broad width.  Is it correlated with their toroid-like or
3-alpha-cluster-like behavior?

In future theoretical work, it will be necessary to include spin-orbit
and other residual interactions so as to obtain the fine-structure  of
the multiplets. 
The knowledge of a better wave function will allow the
evaluation of the toroidal moment of inertia of the Hoyle state, for
comparison with the observed moment of inertia.  How the additional
three-alpha cluster interaction may be included as a residual
interaction in a mean-field dynamics will be of great interest.  The
quantitative evaluation of the overlap amplitude between the toroidal
state and the 3$\alpha$ cluster state and the correction to the
adiabatic energy $\Delta E_{3\alpha}(R) $ arising from the three-alpha
interaction will be of great interest to provide a firmer foundation on
the microscopic description of the toroidal configuration.

%\vspace*{0.2cm}
\centerline{\bf Acknowledgments}

The authors would like to thank Profs.\ Aksel S. Jensen, Larry Zamick,
Yitzak Sharon, Ikuko Hamamoto, Jirina Stone, Hans O. U. Fynbo, Oliver
S. Kirsebom, Soren Sorensen, Howard Holme, and Joe Natowitz for
helpful communications and discussions.  The research was supported in
part by the Division of Nuclear Physics, U.S. Department of Energy
under Contract DE-AC05-00OR22725.

\end{document}